\documentclass[useAMS,usenatbib]{mn2e}
\usepackage{graphicx}

\newcommand{\etal  }{{et al.} }
\newcommand{\msun}{\thinspace M_\odot}  
\newcommand{\vect}[1]{\mbox{\boldmath$#1$}}
\newcommand{\omegac}{\Omega_{\rm c}}
\newcommand{\bzc}{ B_{ zc}}
\newcommand{\amt}{ A_{ \varphi}}
\newcommand{\amz}{ A_{ z}}
\newcommand{\rhoc}{\rho_{\rm c}}

\newcommand{\ob}{\varepsilon_{\rm ob}}
\newcommand{\ar}{\varepsilon_{\rm ar}}
\newcommand{\bb}{ B_{\rm zc}/(8\pi c_s^2 \rho_{\rm c})^{1/2}  }
\newcommand{\ww}{\Omega_{\rm c}/(4\pi  G \rho_{\rm c})^{1/2}}
\newcommand{\cm  }{\,{\rm cm}^{-3} } 
\newcommand{\fig}[1]{Fig.~\ref{fig:#1}}
\newcommand{\eq}[1]{equation~(\ref{eq:#1})}
\newcommand{\jcm}{{\rm cm^2\,s^{-1}}}
\newcommand{\dfrac}[2]{{\displaystyle \frac{#1}{#2}}  }

\title[Collapse and Fragmentation of Molecular Clouds. II. ]
{
Collapse and Fragmentation of Rotating Magnetized Clouds. II. \\
  Binary Formation and Fragmentation of First Cores 
}
\author[M. ~N. ~Machida, T. ~Matsumoto, T. ~Hanawa  and K. ~Tomisaka]
{Masahiro ~N. ~Machida$^{1}$\thanks{E-mail:machida@cfs.chiba-u.ac.jp} 
,Tomoaki ~Matsumoto$^{2}$\thanks{E-mail:matsu@i.hosei.ac.jp} 
,Tomoyuki ~Hanawa$^{1}$\thanks{E-mail:hanawa@cfs.chiba-u.ac.jp} and 
\newauthor  Kohji ~Tomisaka$^{3}$\thanks{E-mail:tomisaka@th.nao.ac.jp} \\
$^{1}$ Center for Frontier Science, Chiba University, Yayoicho 1-33, Inageku, Chiba 263-8522, Japan \\
$^{2}$ Faculty of Humanity and Environment, Hosei University, Fujimi, Chiyoda-ku, Tokyo 102-8160, Japan \\
$^{3}$ National Astronomical Observatory of Japan, Mitaka, Tokyo 181-8588, Japan
}

\begin{document}



\maketitle

\begin{abstract}
Subsequent to Paper I, the evolution and fragmentation of a rotating magnetized cloud are studied with use of 
 three-dimensional MHD nested-grid simulations.  
After the isothermal runaway collapse, an adiabatic gas forms a protostellar first core at the center of the cloud.  
When the isothermal gas is stable for fragmentation in a contracting disk,
 the adiabatic core often breaks into several fragments.  
Conditions for fragmentation and binary formation are studied.  
All the cores which show fragmentations are geometrically thin, as the diameter-to-thickness ratio is larger than 3.  
Two patterns of fragmentation are found.  
(1) When a thin disk is supported by centrifugal force, the disk fragments through a ring configuration (ring fragmentation). 
This is realized in a fast rotating adiabatic core as $\Omega >0.2 \tau_{\rm ff}^{-1}$, where $\Omega$ and $\tau_{\rm ff}$ represent the angular rotation speed and the free-fall time of the core, respectively.  
(2) On the other hand, the disk is deformed to an elongated bar in the isothermal stage for a strongly magnetized or rapidly rotating cloud.
The bar breaks into 2 - 4 fragments (bar fragmentation).  
Even if a disk is thin, the disk dominated by the magnetic force or thermal pressure is stable and forms a single compact body.  
In either ring or bar fragmentation mode, the fragments contract and a pair of outflows are ejected from the vicinities of the compact cores.
The orbital angular momentum is larger than the spin angular momentum  in the ring fragmentation.
On the other hand, fragments often quickly merge in the bar fragmentation, since the orbital angular momentum is smaller than the spin angular momentum in this case.
Comparison with observations is also shown.    

\end{abstract}


\section{Introduction}
Many stars are observed as members of binary or multiple systems \citep[e.g.][]{abt83,duquennoy91}.
The binary frequency increases more in star-forming regions \citep{cohen79,dyck82,joy94,richichi94}.
However, as summarized in \citet{clarke91}, the dynamical evolution of a small number of point-mass stars leads to a system composed of one central binary or triple and many single escapers,  which is a natural outcome of a three-body encounter
 to make a close binary \citep{binney87}.
This does not explain the fact that binaries and multiples are common.
Since the encounters in a cluster of stars or protostars cannot explain all the binary stars
 \citep{kroupa01},
 we should devote attention to the direct binary formation through 
 the fragmentation process in the course of star formation \citep[e.g.][]{bodenheimer00}.

It is known that the molecular cloud collapses isothermally until the gas density reaches $ n \approx 5 \times 10^{10} \cm$  \citep{larson69,tohline82,masunaga00}.
After the gas density exceeds this, gas becomes adiabatic, because the cloud becomes  optically thick against the dust thermal emissions and forms a first core \citep{larson69}.
The continuous accretion increases the mass of the first core and thus the gas density increases steadily in the core.
Simultaneously, the temperature rises steadily.
Finally, molecular hydrogen is dissociated after reaching  $T \simeq 10^3$K and $n \simeq 10^{15} \cm$.
Then the cloud begins to contract again as a result of the endoergic reaction.

The fragmentation of molecular cloud has been investigated by many authors
 (for review, see \citealt{bodenheimer00}).
Majority of researchers have focused on the isothermal clouds.
The initial conditions of the spherical isothermal cloud with uniform density
 and rigid-body rotation are characterized by two parameters:
 thermal-to-gravitational energy ratio $\alpha_0$ and
 rotational-to-gravitational energy ratio $\beta_0$.
The evolution is divided into three types \citep{miyama84,tsuribe99}.
 (1) a cloud with $\alpha_0 \ga 1$ does not contract and oscillates.
 (2) a cloud with $0.5 \la \alpha_0 \la 1$,
 which means the initial state is near the hydrostatic equilibrium 
 from the Virial Theorem,  experiences  run-away collapse. 
The central density increases substantially in a  finite timescale.
In this case, the contracting cloud does not fragment.
(3) for a ``cold'' cloud such as $\alpha_0 \la 0.5$,
 the cloud collapses in the direction of the angular momentum vector and forms a disk.
 The disk is subject to fragmentation
as shown in \citet{bonnell94}, \citet{bonnell94a,bonnell94b} and \citet{whitworth95}
Although \citet{cha03} have reported that even the cloud with 
 $\alpha_0\simeq 0.6$ fragments in the isothermal regime
 if it has a sufficiently strong differential-rotation,
 such a ``warm'' cloud does not fragment unless a ring is formed in the
 isothermal phase.

Since a (nearly) rotation-supported disk is formed in the adiabatic phase,
 the disk fragments if it is sufficiently thin.     
Therefore, even if the gas does not fragment in the isothermal regime, 
 it breaks into fragments after the adiabatic core has formed (the first core).
Actually, \citet{cha03}, and \citet{matsu03} assumed the barotropic
 equation of state for a gas and studied the evolution of a ``warm'' cloud 
 with $\alpha_0=0.6-0.76$.
They found that the thin disk fragments in the adiabatic phase
 even if $\alpha_0$ is chosen for the cloud to experience
 the run-away collapse.
\citet{matsu03} showed that a cloud with $\Omega_0
  t_{\rm ff} \ga 0.05 $ fragments, where $\Omega_0$ and $t_{\rm
  ff}$ denote the initial central angular velocity and the
  free-fall time at the center $(3 \pi / 32 G \rho_0)^{1/2}$,
  respectively.
By comparing different initial rotation laws they found that the central angular velocity is essential to whether the cloud fragments or not.
This shows us that the fragmentation conditions are directly related to
 $\Omega_0 t_{\rm tt}$ rather than $\beta_0$. 
The thickness of the disk is also important in the fragmentation process in the adiabatic stage \citep{matsu03,machida04a}.
Since the Toomore's $Q$ value of the disk is smaller than 1 and 
 the gravitationally most unstable wave-length is much smaller than the disk
 radius, the disk which fragments is gravitationally unstable \citep{matsu03}.

Disk structure is believed to be formed either by the rotation or the magnetic field.
Almost all previous studies focus on the hydrodynamical contraction process.
However, the molecular cloud is magnetized, and  magnetic field lines along the major axis of molecular clouds are often observed \citep[e.g.][]{tamura95,ward00}.
In star-forming regions, the molecular outflows are seen to be ubiquitous \citep[e.g.][]{ohashi96,belloche02}.
\citet{tomisaka98,tomisaka00,tomisaka02} showed that these outflows are driven by the magnetic force in his two-dimensional axisymmetric magnetohydrodynamical (MHD) simulations.
\citet{basu94} pointed out that magnetic braking also plays an important role in a magnetically supported rotating cloud.
Consequently, the magnetic field should play an important role in the star-formation process.

When the cloud has no rotation motion, the magnetic field must have a positive
 effect to form a thin disk \citep[see, e.g.,][]{nakamura95,tomisaka96}.
However, in the rotating cloud, since the angular momentum can be removed
 from the core by magnetic force \citep{basu94,tomisaka00}, 
 the magnetic field may have a negative effect for fragmentation.
To investigate the effect of the magnetic field on fragmentation or
 binary formation, three-dimensional calculation taking account of the magnetic field is needed.

However, there have been few studies of these calculations.
It has not been cleared yet whether the magnetic field promotes fragmentation or suppresses it.
\citet{boss02} studied the evolution of magnetized molecular clouds and he claimed that the magnetic field  promotes fragmentation.
However, in his study, the angular momentum transfer by magnetic tension force is not included,  because the magnetic field is accounted for by an approximate form.
Recently, \cite{hosking04} studied  fragmentation of a magnetized cloud using their MHD SPH code.
They found that fragmentation is prevented by the magnetic field because magnetic braking is so effective in their cloud.
However, they calculated only 4 models with simple initial conditions.
See also Boss (2004) for a counterargument against \citet{hosking04}.
Thus, it has not become clear whether the molecular clouds observed can actually fragment or not.
In our study, having a large parameter range of initial rotation speed and magnetic field strength, it is found that fragmentation is suppressed by the magnetic field, as denoted in \cite{hosking04}, although the magnetic braking is not so effective.

Ambipolar diffusion affects protostellar collapse especially at high density exceeding $ n \simeq 10^{11} \cm$.
However, the diffusion timescale has uncertainty depending on the ionization rate \citep{nakano02}. 
Furthermore, a calculation of ambipolar diffusion requires large computational costs: Hosking \& Whitworth (2004) calculate only a few models, and Boss (2002) adopts a simple approximation
in the treatment of the diffusion. 
We therefore adopt the ideal MHD in order to perform an extensive parameter survey.  
Nakano \etal (2002) shows that the magnetic field is coupled with gas in $n \la 10^{11-12} \cm$, indicating that the assumption of an ideal MHD is valid in the isothermal phase.  
In the adiabatic phase, the number density in the adiabatic core exceeds $\sim 10^{12} \cm$, and the magnetic field begins to decouple. 
Our simulation may therefore overestimate the angular momentum transfer by magnetic field, especially in the dense fragments.

We study the evolution and fragmentation of molecular clouds using full three-dimensional MHD simulations.
In this study, we use a nested grid code, which always maintains sufficient spatial resolution in the central region.
In this simulation, structures in the range of 5 orders of magnitude in spatial extent (e.g. $\sim 10^6$AU $- 10$AU) are resolved, which corresponds to  14 orders of magnitude in the density contrast (e.g. $5\times 10^2 \cm - 10^{17} \cm$).

 We calculated 144 models with different magnetic field strengths, rotation speeds, and  initial amplitudes of non-axisymmetric perturbations.
We have already reported a part of the result and a condition of fragmentation in Machida, Tomisaka \& Matsumoto (2004; hereafter MTM04).  
MTM04 shows that fragmentation occurs if the first core has a large oblateness (radius-height ratio).  
If the disk is thin enough (oblateness $> 4$), a non-axisymmetric mode develops in it after the gas becomes adiabatic.  
When a thick disk or a core is left at the core formation epoch (oblateness $< 4$), however, an axisymmetric core continues to contract.
This seems to indicate single-star formation.  
If the non-axisymmetric perturbation has grown significantly and a bar appears in the isothermal stage, such a bar fragments in the following adiabatic accretion stage.

Subsequent to the companion paper (Machida \etal 2004; hereafter Paper I), we report results of three-dimensional MHD simulations in detail and give a condition for binary formation.  
Paper I has studied the early evolution in which gas behaves isothermally.  
We have found that (1) a disk is formed in a direction perpendicular to the magnetic field and the rotation axis, and (2) the ratio of magnetic flux density to the square root of the gas pressure [$B/(8 \pi \, c_s^2\rho)^{1/2}$] and that of the angular rotation speed to the gravitational free-fall rate [$\Omega/(4\pi G \rho)^{1/2}$], both of which are evaluated at the center of the cloud, are well correlated at the core-formation epoch as 
\begin{equation}
   \frac{\Omega_c^2}{(0.2)^2 \times 4 \pi G \rho_c} + \frac{B_{\rm zc}^2}{(0.36)^2 \times 8 \pi c_s^2 \rho_c} =1.
\label{eq:UL}
\end{equation}
We evaluated the amplitude of the nonaxisymmetricity with the axis ratio viewed from the top at the end of the isothermal phase.
This controls the subsequent evolution.
We calculated  cloud evolution in the density range from $n = 5 \times 10^2 \cm$ to $\approx 10^{17} \cm$.
In this paper, we present the evolution of the adiabatic accretion stage ($n > 5\times 10^{10} \cm$), while that of the isothermal collapse stage ($n \le 5 \times 10^{10} \cm$) was shown in Paper I.
Model and numerical method are given in \S 2, and the results of our calculation are presented in \S 3.
 In \S 4, we discuss the fragmentation conditions and the angular momentum redistribution between orbital and spin angular momenta after the fragmentation.

\section{Model and Numerical Method}

The initial model and numerical method are the same as those of MTM04 and Paper I.
Here we describe them briefly.  
We assume that gas obeys the ideal MHD equations (eqs.[I.1]--[I.4]) for simplicity, where equation [I.1] represents equation (1) of Paper I.  
Interstellar gas behaves isothermally for $n \la 5\times 10^{10} \cm$ and adiabatically with the specific heat ratio of $\gamma=7/5$ for $n \ga 5\times 10^{10} \cm$\citep{larson69,tohline82,masunaga00}.  
Thus, we assume a barotropic equation of state as 
\begin{equation}
P = c_s^2 \rho \left[
1 + \left( \dfrac{n}{n_{\rm cri}}   \right)^{2/5} 
\right],
\end{equation}
where $c_s$ denotes the isothermal sound speed and $\rho_{\rm cri} = 1.9 \times 10^{-13}\,$g$\cm$ or $n_{\rm cri} = 5 \times 10^{10} \cm$ \citep{bonnell94b,bate02}.
As in MTM04 and Paper I, a filamentary cloud of infinite length is assumed as the initial state, which is in a hydrostatic balance radially.  
To this hydrostatic distribution we added axisymmetric density perturbation $\delta \rho_z(z)$ as well as non-axisymmetric density $\delta \rho_\varphi(r,\varphi)$ and magnetic flux density perturbations $\delta \vect{B}_\varphi(r,\varphi)$ as equations [I.7] and [I.9].  

We chose to describe the amplitude of axisymmetric density perturbation are 
proportional to the unperturbed density with the amplitude fixed at 10\%.  
The initial model is characterized by three nondimensional parameters: 
the magnetic-to-thermal pressure ratio,
\begin{equation}
\alpha =  B_{\rm zc,0}^2 / (4\pi \rho_{\rm c,0} c_{\rm s,0}^2) ,
\label{eq:alpha}
\end{equation}
the angular rotation velocity normalized by the free-fall timescale,
\begin{equation}
\omega = \Omega_{\rm c,0}/\left( 4 \pi  G \rho_{\rm c,0} \right)^{1/2} ,
\label{eq:omega}
\end{equation}
and the initial amplitude of the non-axisymmetric perturbation, $ \amt $.
The former two specify the equilibrium model, while the latter does the perturbations.
We made 144 models by combining the above parameters and calculated them with the 3D nested grid code, as we did in MTM04 and Paper I.
We assume that the cloud has a central density of $n_0 = 5\times 10^2 \cm$ initially, except for some models in which the cloud has a central density of $5\times 10^4 \cm$ and $5\times 10^6 \cm$. 
In the following sections, we mention the evolution of the clouds only with $n_0 = 5 \times 10^2 \cm$ and discuss its dependency on the initial density in \S 4.1.

\section{Results}
\label{sec:results}
When the central density exceeds $n_{\rm c}  \simeq 5 \times 10^{10} \cm$, the gas temperature begins to rise, which supports the gas and forms a static core (the first core) as denoted in \citet{larson69}.
When the first core rotates or has a magnetic field, various substructures (c.f. core, disk, ring, and bar) appear.
That is, after the gas becomes adiabatic,  fragmentation may occur as shown in MTM04 (see also \citealt{matsu03} for a hydrodynamical case).
In this section, we discuss  cloud evolution in the adiabatic accretion phase ($n_{\rm c}> 5 \times 10^{10} \cm$), devoting particular attention to the fragmentation patterns.
As shown in MTM04, the adiabatic core shows three characteristically different types of evolution.
In the case with a strong magnetic field, magnetic braking works and the angular momentum is efficiently removed from the core.
This results in a compact adiabatic core.
From a fast rotating cloud, a thin disk appears and this evolves into a ring.
The ring breaks through a  non-axisymmetric spiral mode and leads to binary fragments, which is called the `ring fragmentation' mode.
The last case, in which a large amplitude of the initial nonaxisymetric perturbation forms a bar,  eventually fragments into two or several pieces.
The fragments contract and form compact cores.
This is called the `bar fragmentation' mode.

We show these three models, the core, ring fragmentation and bar fragmentation models in greater detail in the following sections.
The model parameters and physical quantities are summarized in Table~\ref{table:1}.
In this table, the models named  BL (core model), CS (ring fragmentation model), and DL (bar fragmentation model) are the same as those in Paper I.
The ratio of the total to the magnetically critical mass, $M / M_{\rm B,cri}$ ( $M_{\rm B,cri} \equiv B_{\rm z,c0}/2 \pi G^{1/2}$), is also described in this table.
All models are magnetically supercritical cloud at initial, because $M / M_{\rm B,cri}$ exceeds unity.
The evolution of these clouds in the adiabatic accretion phase ($n_c \ge 5 \times 10^{10} \cm$) is shown in this paper, while that in the isothermal collapse phase ($n_c < 5 \times 10^{10} \cm$) is described in Paper I.

\subsection{Core Model}
\begin{figure}
\begin{center}
\includegraphics[width=150mm]{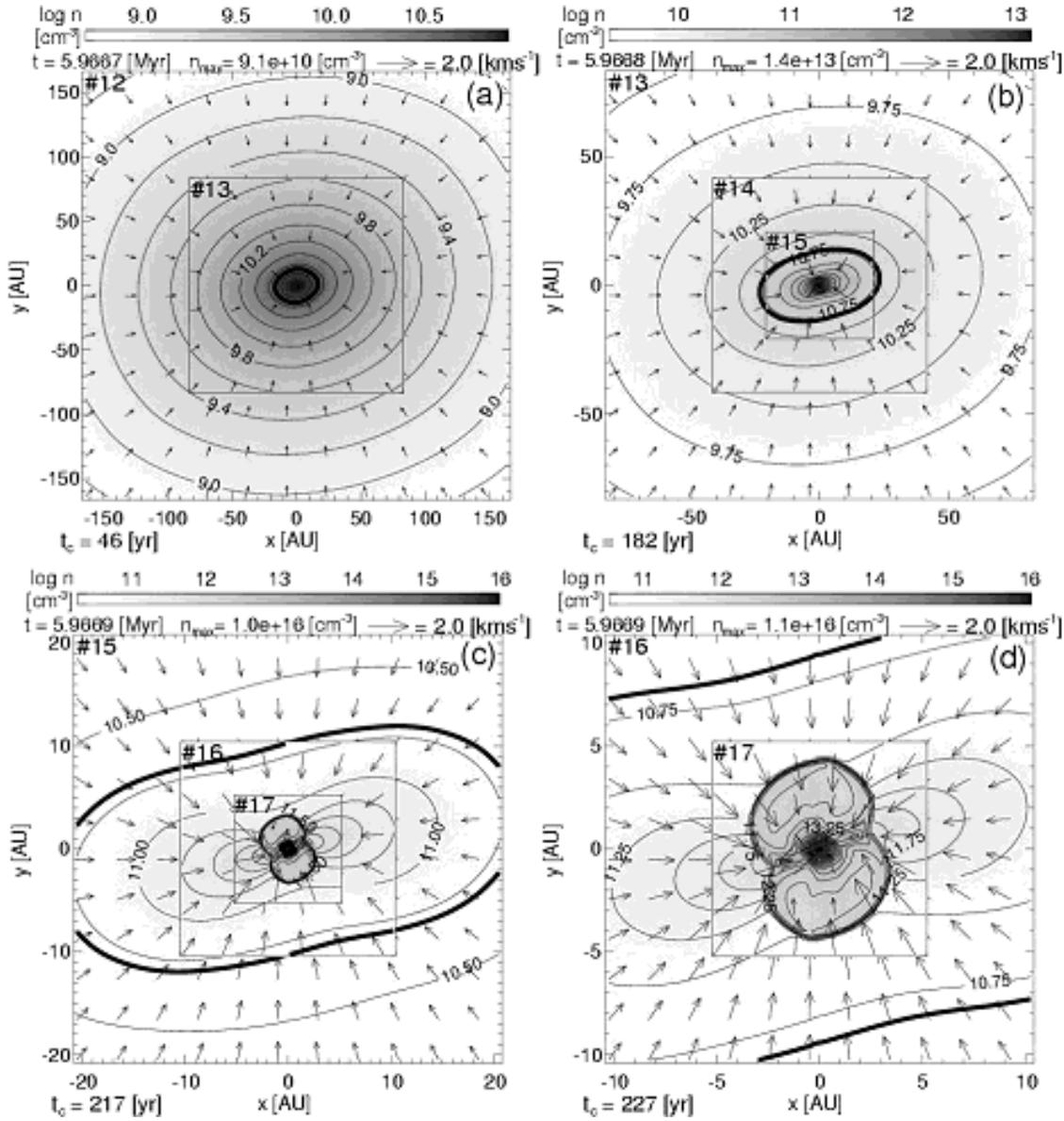}
\caption{
The density (false color and contours) and velocity distributions (arrows) for Model BL [($\alpha$, $\omega$, $\amt$)=(0.1, 0.01, 0.2)] are plotted on the $z = 0$ plane at $t_c=46$ yr (a) , 182 yr (b), 217 yr (c), and 227 yr (d) after core formation epoch of $n_c = 5 \times 10^{10} \cm$. The time elapsed from the beginning of the calculation, the maximum density and the scale of the velocity vectors are displayed in the upper area of each panel.
The time after the adiabatic core formation is shown in the lower region of each panel.
Levels of grid are shown in the upper left corner of the boundaries of subgrids.
The thick line denotes the adiabatic core, namely, it means the contour of $ n = 5\times 10^{10} \cm$.
}
\label{fig:core1}
\end{center}
\end{figure}

\begin{figure}
\begin{center}
\includegraphics[width=120mm]{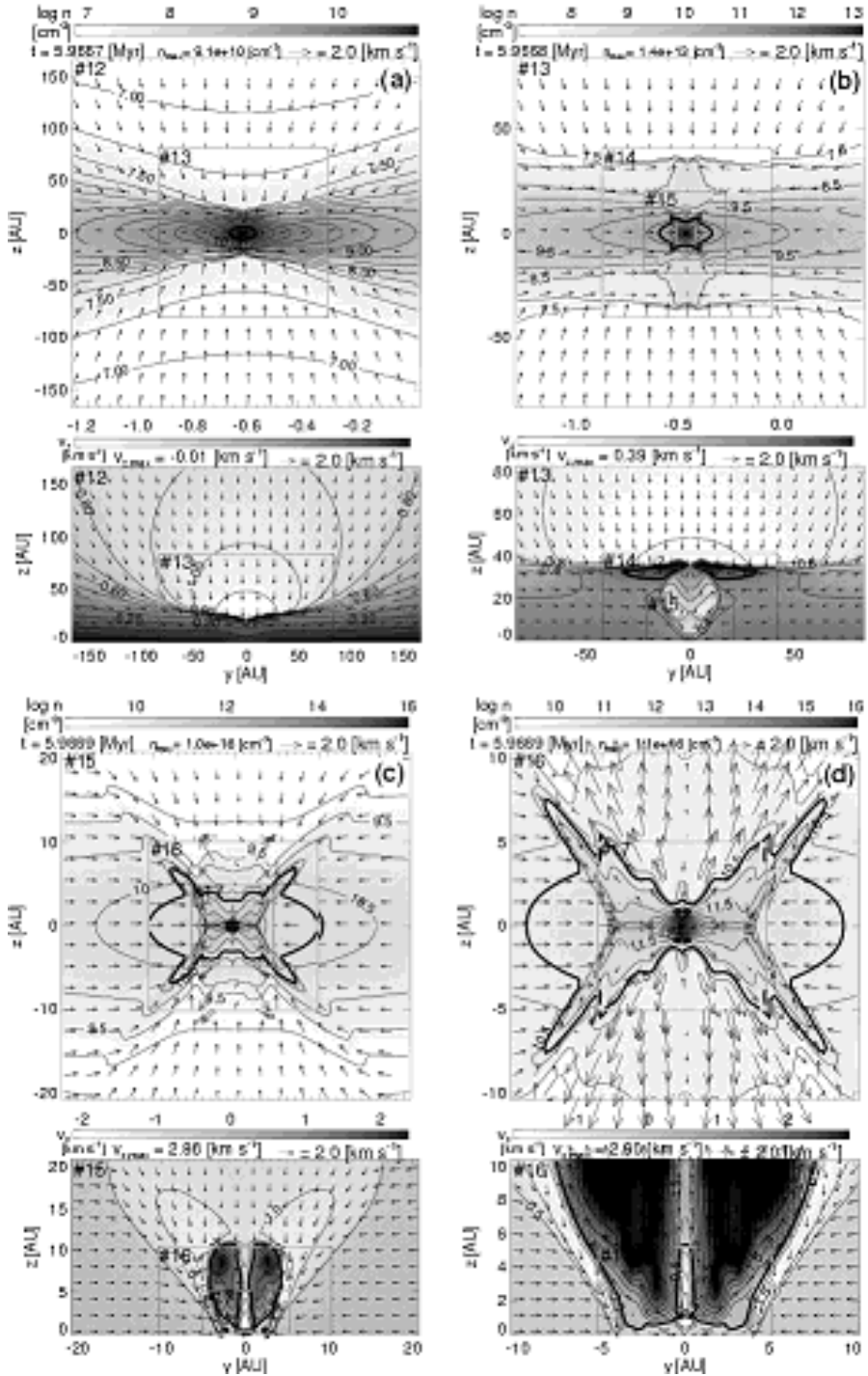}
\caption{
In each upper panel, the density (false color and contours) and velocity distributions (arrows) on $x = 0$ plane are plotted for model BL [($\alpha$, $\omega$, $\amt$)=(0.1, 0.01, 0.2)]  at the same epochs of \fig{core1}.
Thick lines denote the adiabatic core, namely, they mean the contour of $n = 5 \times 10^{10} \cm$.
In each lower panel, $v_{\rm z}$ (false color and contours) and  velocity vectors ($v_y$, $v_z$: arrows) on the $x$=0 plane are plotted.
The maximum speed in the $z$-direction ($v_{\rm z,max}$) is also shown in the upper part of the lower panels.
Thick lines denote the boundary of $v_{\rm z}=0$.
Inside the thick line, gas has an upward velocity ($v_{\rm z}>0$).
}
\label{fig:core2}
\end{center}
\end{figure}

First, we show the typical evolution of the `core' model (model BL) in the adiabatic accretion phase.
A cloud with a weak magnetic field ($\alpha <0.1$ ) and slow rotation ($\omega <0.1$) forms a thick disk or an oblate spheroid at the end of the isothermal collapse phase ($n_{\rm c} \la 5 \times 10^{10} \cm$).
In these models, the central density continues to increase.
In the central region, a dense and compact core appears, but this evolves without any indications of fragmentation during the adiabatic accretion phase ($n_{\rm c} \ge 5 \times 10^{10} \cm$).
Figs.~\ref{fig:core1} and \ref{fig:core2} show the evolution of model BL ($\alpha$, $\omega$, $\amt$)=(0.1, 0.01, 0.2).
We calculate this model from the maximum density of $5\times 10^2 \cm$ to $5\times 10^{16} \cm$.
However, in this section, we focus on the evolution only in the adiabatic accretion phase [see \S 4.2 of Paper I for the isothermal evolution].
At the beginning of the adiabatic accretion stage, the central oblateness and axis ratio reach, respectively, $\ob = 5.3$ and $\ar =0.23$.
\footnote{
Let the length of principal axes $a_1 \geq a_2$ in the $x-y$ plane and $a_z$ in the $z-$direction.
The oblateness ($\ob$) and axis ratio ($\ar$) are defined $\sqrt{a_1\, a_2}/a_3$ and $a_1/a_2 -1$ [see \S 4.1 of Paper I ].
}
In Figs.~\ref{fig:core1} and \ref{fig:core2}, panels (a) through (d) are snapshots at $t_c =$ 46 yr (a), 182 yr (b), 217 yr (c) and 227 yr (d) after the core formation epoch, respectively.
We define the core formation epoch ($t_{\rm c}=0$) as the epoch in which the central density reaches $n_c = 5 \times 10^{10} \cm$.
The thick lines in Fig.~\ref{fig:core1}, which correspond to the contour lines of $n\, =\, 5 \times 10^{10} \cm$, indicate the adiabatic core. 
In the lower panels of \fig{core2}, other thick lines are drawn to indicate the outflow region with $v_z>0$.

\fig{core1} (a) shows that the adiabatic core (inner thick line) has an almost axisymmetric shape and the diameter of the core is about $45$ AU on $z$=0 plane.
Gas inflows radially and the inflow speed reaches $v_r = -0.54$ km\,s$^{-1}$ on the $z=$0 plane, while the gas has a low rotation speed of $v_{\varphi}= 0.06$ km\,s$^{-1}$.
At the same time, the gas forms a thick disk with a height of 36 AU as shown in the upper panel of \fig{core2} (a).
It shows that gas flows vertically outside the disk, while it flows radially inside the disk [the lower panel of \fig{core2}(a)].

The central density increases  from $9.1 \times 10^{10} \cm$ [\fig{core1} (a)] to $1.4 \times 10^{13} \cm$ [\fig{core1} (b)].
The first core changed shape and become ellipsoid, as shown in \fig{core1} (b).
This shows that the non-axisymmetry grows after the gas becomes adiabatic.
\fig{core2} (b) ($x$=0 cut) shows that this ellipsoidal core thickens and the vertical-to-horizontal axis ratio of the adiabatic core increases from 0.2 [\fig{core2} (a)]  to 0.4 because the gas temperature and, accordingly, the sound speed increases in this phase.
This panel shows two shocks.
The outer shock front is seen at $z \simeq \pm 40$AU, while the inner shock is seen in the region  of $ \vert z \vert \la 40$AU and $ \vert y \vert \la 15$AU.
\fig{core2}(b) shows that the gas inside  the inner shock has a higher infall velocity ($v_z\simeq -1.0\, $km~s$^{-1}$) than that outside the inner shock ($v_z\simeq -0.2\, $km~s$^{-1}$).

Fig.~\ref{fig:core1}(c) and (d) show that the adiabatic core is composed of a small core and a disk.
Figs.~\ref{fig:core1}(c) and (d) indicate that the outer boundary of the adiabatic core appears almost the same as  that of \fig{core1}(b) [Panels (c) and (d) are 4 and 8 times more magnified than that of panel (b), respectively].
At the stage of \fig{core1}(d), the central density reaches $1.1 \times 10^{16} \cm$ and a dense compact core with a steep density gradient has formed at the center.
A disk with density in the range from $ 2 \times 10^{11} \cm$ to $5 \times 10^{13}\cm$ surrounds the core.

\begin{figure}
\begin{center}
\includegraphics[width=150mm]{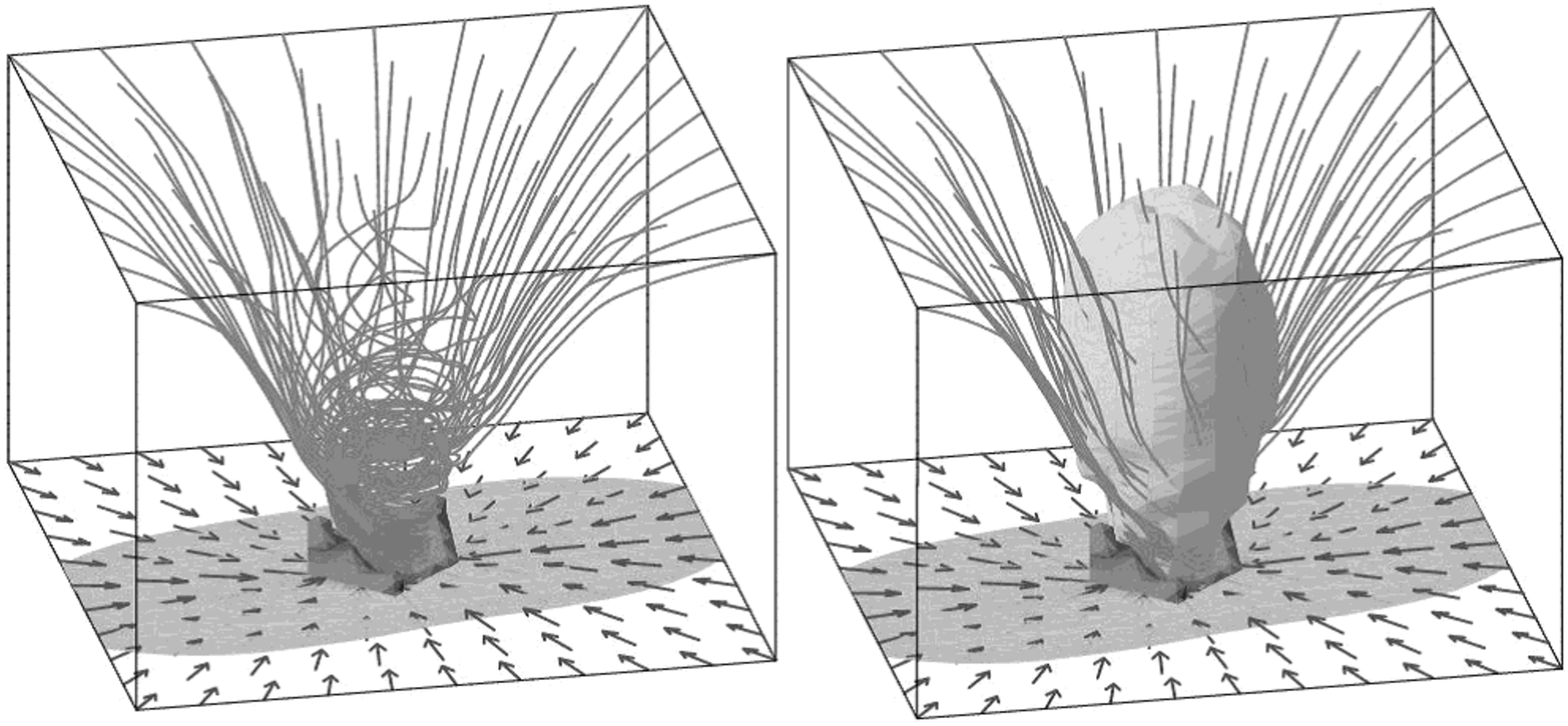}
\caption{
The magnetic field lines and outflow region are plotted for model BL [($\alpha$, $\omega$, $\amt$)=(0.1, 0.01, 0.2)] at the same epoch of \fig{core1} (d).
Magnetic field lines (stream lines), adiabatic core (isodensity surface) and velocity vectors (arrows) on the $z$=0 plane are plotted in the left panel.
In the right panel, the outflow regions (isovelocity surface of $v_z = 0$) are also plotted beside the magnetic field lines, the adiabatic core and the velocity vectors.
The box size is 40 AU.
}
\label{fig:core3}
\end{center}
\end{figure}

As shown by \citet{tomisaka98,tomisaka02}, magnetic field lines are tightly twisted after the adiabatic core formation.
\fig{core3} left panel shows the magnetic field lines (streamlines) at the same stage of \fig{core1} (d).
This shows the magnetic field lines are tightly twisted by the rotating adiabatic core (red isodensity surface).
The magnetic field lines begin to twist because the collapse timescale exceeds the rotational timescale in the adiabatic accretion phase.
The twisted field transfers the angular momentum of the adiabatic core to the outflow, and the outflow emerges from the core and brings the angular momentum from the core.
In this model, the angular momentum is extracted through the gravitational 
and pressure torque since the adiabatic core is appreciable elongated. 
Thus, it is difficult to evaluate the angular momentum extraction 
due to the outflow separately.
However, we have confirmed that the angular momentum is largely removed 
by the outflow in the model of which parameters are the same as model BL
except for $ \amt $ = 0.01.  The angular momentum extraction is mostly
due to the outflow since the core is nearly axisymmetric in the model
of $ \amt $ = 0.01.
For this reason, the angular momentum of the adiabatic core continues to be removed  by the magnetic tension, and thus collapse continues.

In the lower panels of \fig{core2}(c) and (d), a strong outflow is seen, which is driven by the spinning of the adiabatic core.
The outflow reaches $z \simeq$ 10 AU with $v_{\rm z,out} \simeq$ 2.86 km\,s$^{-1}$ of the maximum outflow speed at the stage of \fig{core2} (c), and extends further to $z \simeq 18 $AU with $v_{\rm z,out} \simeq 2.90$ km~s$^{-1}$ at the stage of \fig{core2} (d).
Fig.2(d) upper panel indicates a typical structure of the adiabatic core (thick line) from which outflow is ejected.  
There is a horn which extends from the oblate spheroidal adiabatic core. 
This separates the inflowing and outflowing gases [lower panel of Fig.2 (d)].  
Inside the horn including the rotation axis, gas is outflowing with a speed up to $v_z\simeq 2.9 {\rm km\,s}^{-1}$ and the outside the horn near the disk mid-plane gas is inflowing radially with $v_r\simeq -0.5 {\rm km\,s}^{-1}$.

This configuration of a compact core with a disk is seen also in other models forming a single core in the
 adiabatic phase.
Some hydrodynamical models with no magnetic field $\alpha=0$ ($\omega < 0.02$) belong to this core-type solution.
However, this core $+$ disk configuration does not appear in the models without magnetic field. 
In such models of $\alpha=0$, the adiabatic core is a simple thin disk, and this adiabatic disk gradually contracts.
This seems to indicate that the compact core in the adiabatic outer core or disk is composed of the gas 
 from which angular momentum is extracted by magnetic braking.
The magnetic braking supresses the bar mode instability which 
took place in Bate (1998) and Matsumoto \& Hanawa (2003), who 
assumed no magnetic field.

\subsection{Ring Fragmentation Model}
\label{sec:ringf}
In this section, we show the second typical evolution, e.g. `ring fragmentation' model (model CS) in the adiabatic accretion phase.
This type of fragmentation is also seen in \citet{norman78,tohline80,boss80}. 
When the bar mode does not evolve so much in the isothermal collapse phase (see \S 4.3 of Paper I), the central gas forms an axisymmetric disk at the beginning of the adiabatic accretion phase.
In the later adiabatic accretion phase, the disk is deformed to a ring, which is defined in this paper, as  a structure where the central density is lower than the surroundings.

\begin{figure}
\begin{center}
\includegraphics[width=150mm]{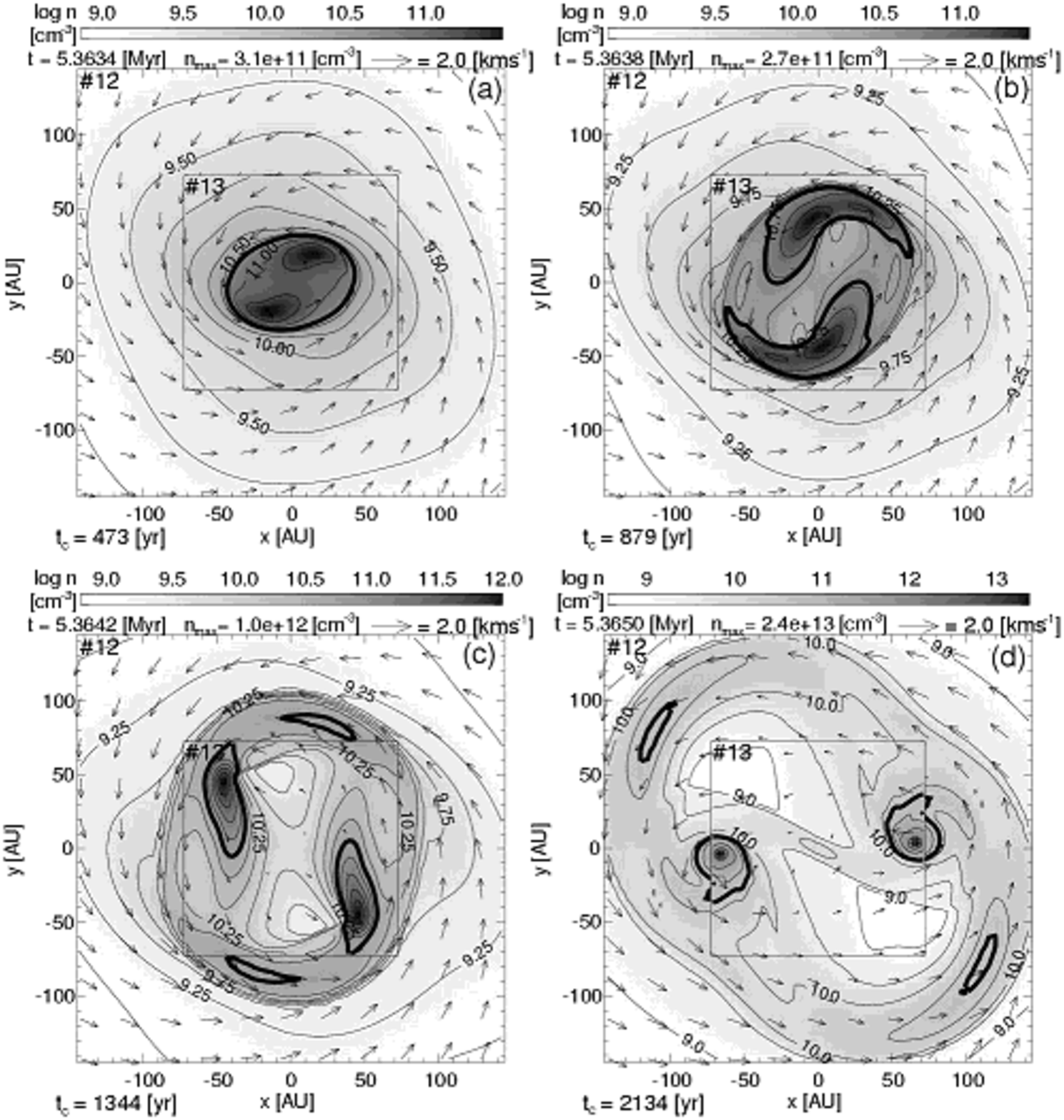}
\caption{
The density (false color and contours) and velocity distributions (arrows) for Model CS [($\alpha$, $\omega$, $\amt$)=(0.01, 0.5, 0.01)] are plotted on the $z = 0$ plane at $t_c=473$ yr (a) ,  879 yr (b), 1344 yr (c), and 2134 yr (d).
The contours, arrows, and notation have the same meaning as in \fig{core1}.
}
\label{fig:ring1}
\end{center}

\end{figure}

\begin{figure}
\begin{center}
\includegraphics[width=130mm]{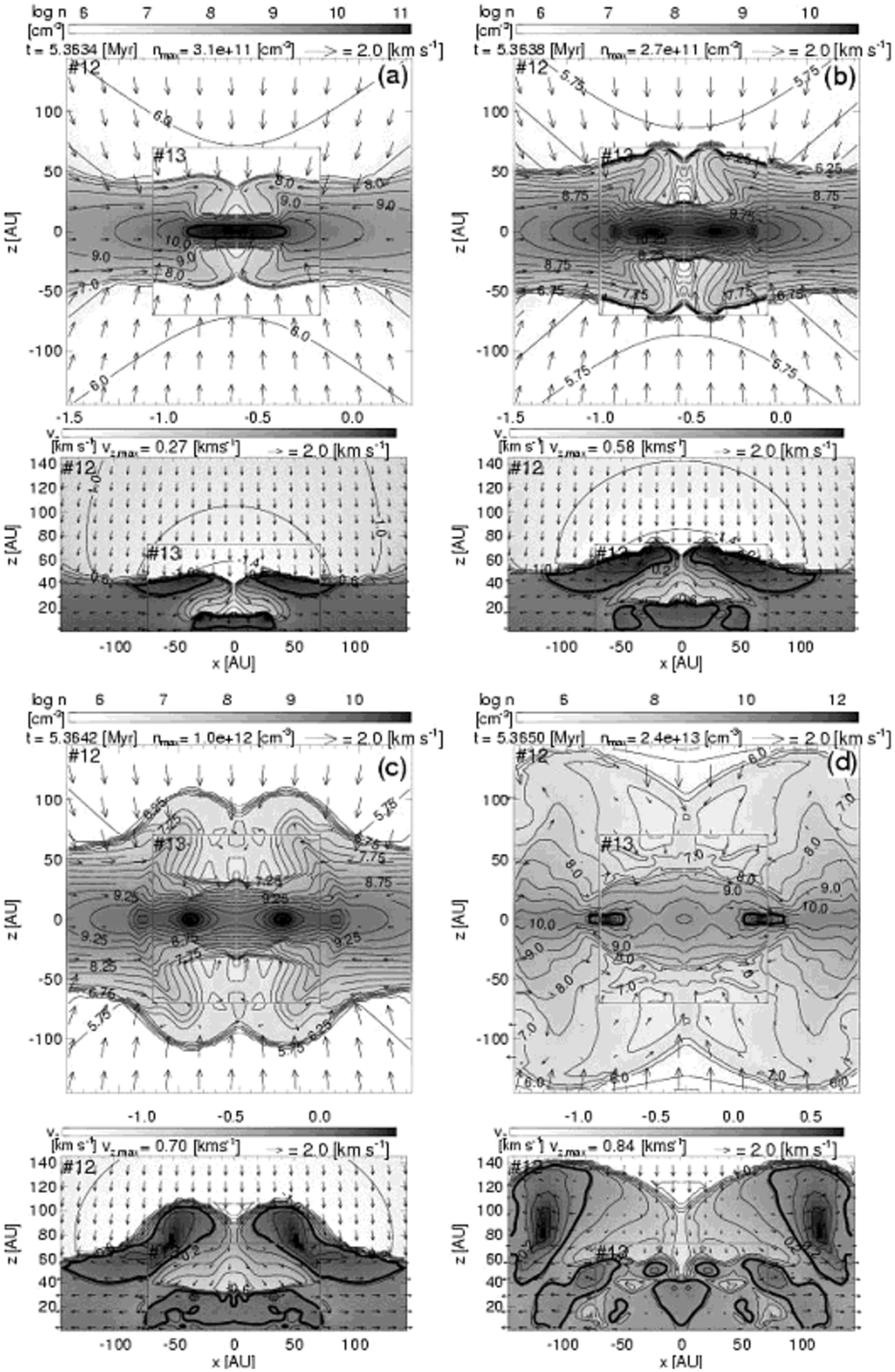}
\caption{
In each upper panel, the density (false color and contours) and velocity distributions (arrows) on $y = 0$ plane are plotted for CS [($\alpha$, $\omega$, $\amt$) = (0.01, 0.5, 0.01)] at the same epochs of \fig{ring1}.
In each lower panel, $v_{\rm z}$ (false color and contours) and  velocity vectors ($v_y$, $v_z$: arrows) on the $x$=0 plane are plotted.
The contours, arrows, and notations have the same meaning as in \fig{core1}.
}
\label{fig:ring2}
\end{center}
\end{figure}

Figs.~\ref{fig:ring1} and \ref{fig:ring2} are the same as Figs.~\ref{fig:core1} and \ref{fig:core2} but for the `ring fragmentation' model CS [($\alpha$, $\omega$, $\amt$)=(0.01, 0.5, 0.01)]. 
In consequence to \S 4.3 of Paper I for the isothermal evolution, we focus on the evolution 
in the adiabatic accretion phase.
In this model, the gas interior to the isodensity surface of 1/10 of the maximum density has an oblateness of 5.1  and the axis ratio of 0.2 (the major-to-minor axis ratio on the $x$-$y$ plane is equal to 1.2) at the core formation epoch.
This means that a thin and almost axisymmetric disk is formed in the isothermal collapse phase. 
Panels (a) through (d) are snapshots at $t_c=$473 yr, 879 yr, 1344 yr, and 2134 yr from the core formation
 in  Figs.~\ref{fig:ring1} and \ref{fig:ring2}.

473 yrs after the core formation epoch, the density peak moves from the center to the  periphery ($r \simeq 20$ AU); that is, the adiabatic core changes shape and becomes a ring.
At the same time, an $m = 2$  non-axisymmetric mode grows and the core begins to fragment.
In \fig{ring1} (a), we can see two fragments whose density peaks are located at ($x$, $y$) = (15 AU, 20 AU) and ($-15$ AU, $-20$ AU).
The axis ratio of the adiabatic core is $\ar = 0.5$ (major-to-minor axis ratio $\simeq$ 1.5).
In \fig{ring2} (a), we find two nested disks, in which the outer disk extends in the region $-50$AU $\la z \la 50$AU, while the inner adiabatic disk (inside a thick line) extends near $-5$ AU $\la z \la 5$ AU and $-40$ AU $\la x \la 40$ AU. 
Both the inner and outer disks are bound by the shock.
The outer shock, which is similar to that seen in the upper panel of \fig{core2}(b), has occurred in the isothermal collapse phase (see \S 4.3 of Paper I).
Gas falls vertically and has a maximum infall velocity of 1.5 km\,s$^{-1}$  just outside the outer shock front ($z \simeq \pm 50$ AU).
Passing through the shock front, gas decelerates to $v_z \simeq -0.1 {\rm km\,s^{-1}}$ of the infall velocity.
However, the infalling gas near the $z$-axis accelerates to 1.2 km~s$^{-1}$, similar to the core model of \fig{core2}.

The low-contrast spiral structure seen in \fig{ring1}(b) grows with time.
Moreover, the adiabatic core indicated by the thick line itself evolves into two fragments.
In this process, a pair of adiabatic cores form.
The peak density  of the fragments is as large as $2.7 \times 10^{11} \cm$, and the peaks are located at ($x$, $y$) $\simeq$ ($5$ AU, $-40$ AU) and ($-5$ AU, $40$ AU).
Comparing the locations of the density peaks  of \fig{ring1} (a) and (b), we can see the fragments revolve counter-clockwise around the center and simultaneously move outward.
On the other hand, the central density decreases to $n_c \simeq 2\times 10^{10} \cm$, and in this range, gas behaves isothermally and no longer adiabatically.
Consequently, a ring-like or two-arm spiral structure is established by this stage.

The fragments and ring structure at $t_c$ = 1344 yr are shown in \fig{ring1} (c).
The density peaks of the fragments are located at ($x$, $y$) $\simeq$ ($50$ AU, $-50$ AU) and ($-50$ AU, $50$ AU), and the maximum density increases up to $\sim 10^{12} \cm$, while the gas density at the center decreases to $ \sim 5 \times 10^9 \cm$.
Thus, there is a density contrast of 1000:1 between the ring or the spiral fragments and the center. 
This panel shows  four adiabatic cores.
The lower density adiabatic cores ($n \simeq 5\times 10^{10} \cm$) near ($x$,$y$) $\simeq$ (20 AU, 80 AU) and ($-$20 AU, $-$80 AU) are formed in the remnant of the ring and at the trailing end of the dense adiabatic cores.
\fig{ring2} (c) upper panel shows that the shock front expands further vertically.

The fragments are seen near ($x$, $y$) $\simeq$ ($65$ AU, $5$ AU) and ($-65$ AU, $-5$ AU) in \fig{ring1} (d).
At this stage, the ring structure of $r \, \simeq\, 100$AU $-170$AU is moving outward.
This expansion of the ring seems due to the fact that the accreting matter has 
 a larger specific angular momentum than does the ring and thus the accretion increases the specific angular
 momentum of the ring.
The fragments continue to move outwardly and at the same time become more compact and dense,
 while the density of the ring gradually decreases. 
Between \fig{ring1} (a) and (d), the fragments revolve $\simeq 180$ degrees around their orbits, although they revolve $\simeq$ 720 degrees until this calculation ends.
The fragments around the orbits revolve rapidly around each other ($v_{\rm \varphi,orb} = 1.6 $km~s$^{-1}$), while they rotate themselves slowly ($v_{\rm \varphi,spn} = 0.5 $km~s$^{-1}$).
It should be noticed that the angular rotation speeds of the orbital motion $\Omega_{\rm orb}$  are 1.5 times higher than those of the spin motion  $\Omega_{\rm spn}$.

Now let's move on to the relation between fragmentation and the outflow.
In the lower panels of \fig{ring2} (a) - (d), thick lines indicate the isovelocity contours of $v_z=0$, which mean the boundaries between the infall and outflow  gases.
Gas is outflowing inside the thick lines.
These panels show that the outflow region expands and the outflow accelerates with time.
In the lower panel of \fig{ring2} (a), we can see two outflow regions, one near the shock front at $z\simeq 50$ AU and the other near the adiabatic core ($z \ga 10$ AU).
The maximum outflow speed is at most $v_{\rm z,out} \simeq 0.27$ km\,s$^{-1}$ at this time.
At the stage of \fig{ring2} (b), the maximum outflow speed reaches $v_{z,{\rm out}} \simeq $
 0.58 km\,s$^{-1}$, while the maximum infall speed reaches  $\simeq$ 1.5 km\,s$^{-1}$.
As the outflow region expands, the inner accretion shock front moves in the $z$-direction from $z \simeq \pm 50$AU [\fig{ring2}(a)] to $z \simeq \pm 70$AU [\fig{ring2}(b)].
As the adiabatic cores (fragments) move radially outward, the outflow region expands radially, too.
The lower panel of \fig{ring2}(c) shows the outflow is further accelerated and its speed reaches $v_{\rm z,out} \simeq 0.7$ km\,s$^{-1}$, while the infall velocity remains $\simeq$ 1.5 km\,s$^{-1}$ at maximum. 
Driven by a strong outflow, the shock fronts move upward to $z \simeq \pm 100$ AU.

\begin{figure}
\begin{center}
\includegraphics[width=150mm]{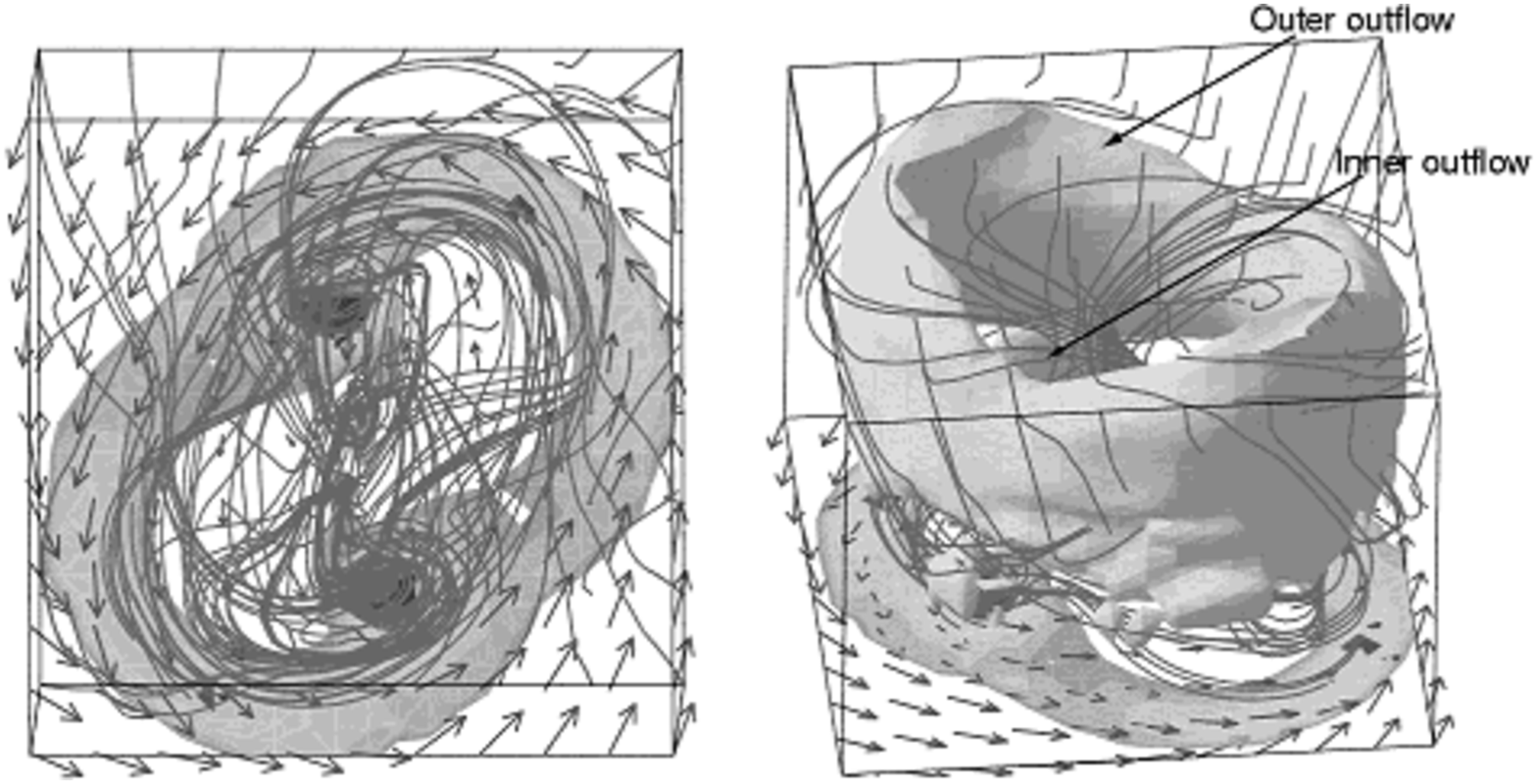}
\caption{
The magnetic field lines and outflow region are plotted for model CS [($\alpha$, $\omega$, $\amt$) = (0.01, 0.5, 0.01)] at the same epoch of \fig{ring1} (d).
Magnetic field lines (stream lines), adiabatic core (isodensity surface) and velocity vectors (arrows) on $z$=0 plane are plotted in the left panel.
In the right panel, the outflow regions (isovelocity surface of $v_z = 0$) are also plotted beside the magnetic field lines, the adiabatic core and the velocity vectors.
The box size is 180 AU.
}
\label{fig:ring3}
\end{center}
\end{figure}

As the ring expands, the outflow region also expands radially.
\fig{ring2}(d) shows that two types of outflow exit.
That is, the outer outflows are connected to the ring structure and the inner ones are connected to the fragments.
The outer outflows have a maximum speed of 0.84 km\,s$^{-1}$ near ($x$,$z$) $\simeq$ ($125$ AU, 90 AU) and ($-125$ AU, 90 AU). 
The inner outflows are seen just above the spinning fragments near ($x$, $z$) $\simeq$ (80 AU, 50 AU), and ($-80$ AU, $50$ AU) and have a complicated shape.
In \fig{ring3}, we can see the relation of the magnetic field lines and the outflow regions more clearly.
In this figure, the magnetic field lines (left panel) and the distribution of outflow regions (right panel) are shown for the $l=12$ subgrid (the box size is equal to 180 AU) from a bird's eye view for the same epoch of \fig{ring1} (d).
This shows that almost all the magnetic field lines near the adiabatic core ($r \la $68 AU)
 are launched from the two fragments.
This is natural, since at least 75 \% of the mass in the adiabatic core is concentrated in these fragments.    
The left panel shows that the magnetic field lines (streamlines) are tangled in the region
 above the remnant of the ring, which is indicated by density contours on $z = 0$ plane, and 
 are strongly twisted by the orbital azimuthal motion of the ring.
\fig{ring3} also shows that the two types of outflow are driven by the rotational motion of the ring and the spin of the fragments, respectively.
The outer outflow has a cylindrical shape and surrounds the smaller-scale outflows.
The outflows connected to the fragments occupy the region interior to the ring-connected outer outflow, and are accelerated with core contraction similar to the outflow appearing in the core model (\fig{core3}).
However, the maximum speed of the outflow is slower than that attained in the bar fragmentation case.
This may be related to the fact the angular rotation speed of the spinning of the fragments,
 $\Omega_{\rm spin}$, is at most comparable to that of the orbital motion, $\Omega_{\rm orbit}$, in the ring.    
Other models which show the ring fragmentation indicate similar evolutions: fragments move outwardly  and two types of outflow co-exist.

\subsection{Bar Fragmentation Model}
\label{sec:barf}
\begin{figure}
\begin{center}
\includegraphics[width=150mm]{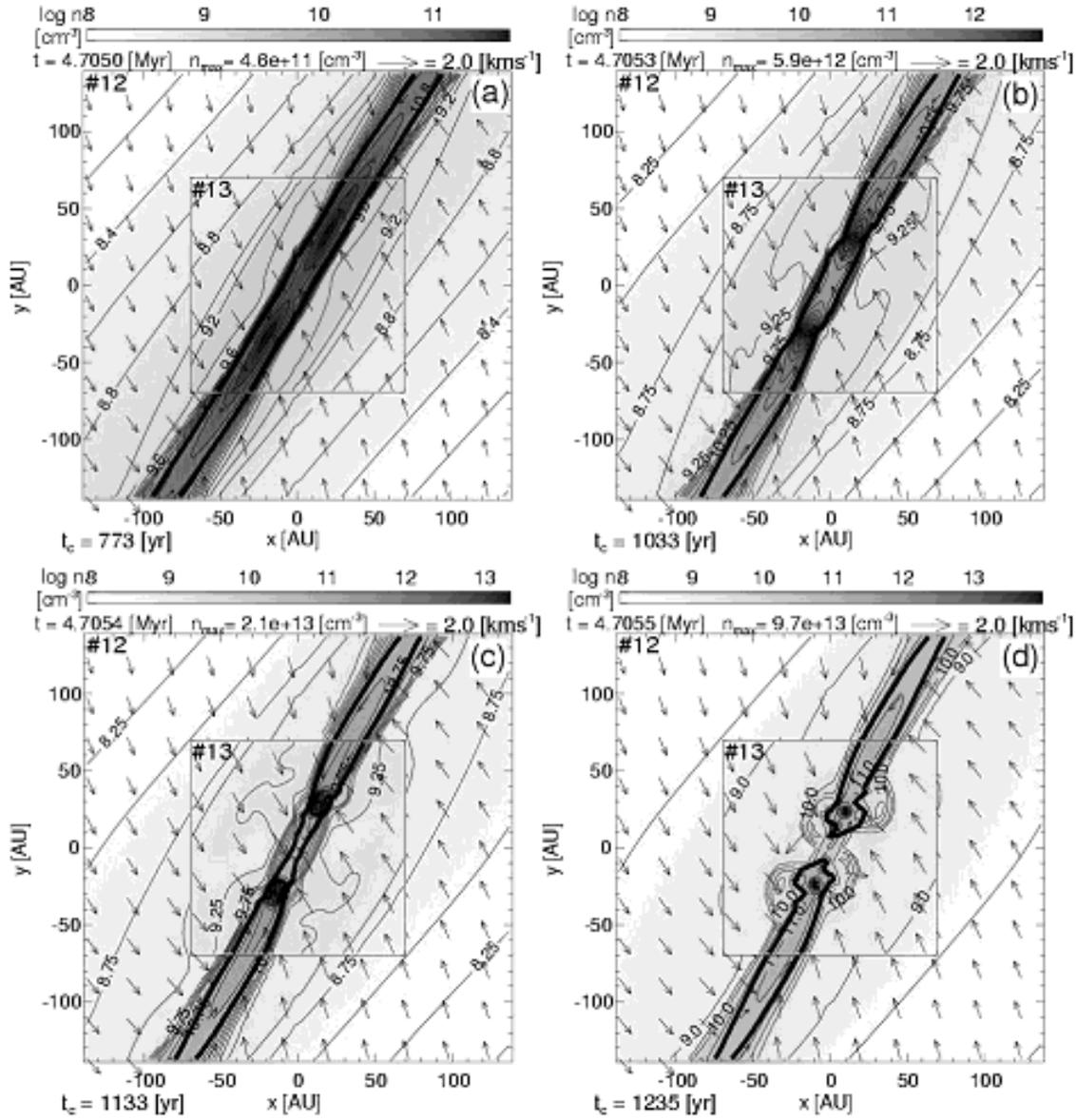}
\caption{
The density (false color and contours) and velocity distributions (arrows) for Model DL [($\alpha, \omega, \amt$) = (1, 0.5, 0.2)] are plotted on the $z = 0$ plane at $t_c=773$ yr (a) ,  1033 yr (b), 1133 yr (c), and 1235 yr (d).
The contours, arrows, and notations have the same meaning as in \fig{core1}.
}
\label{fig:bar1}
\end{center}
\end{figure}

\begin{figure}
\begin{center}
\includegraphics[width=140mm]{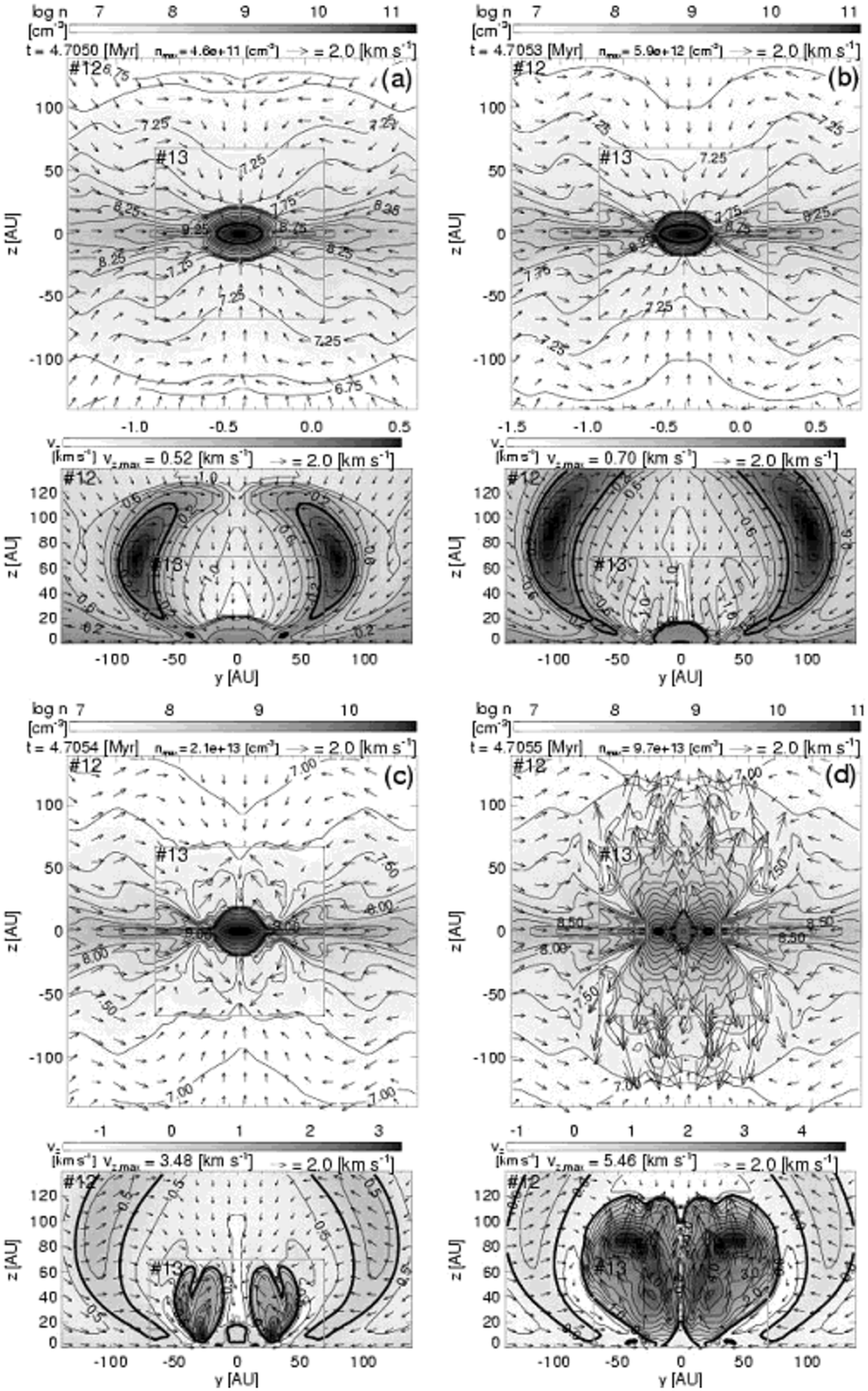}
\caption{
In each upper panel, the density (false color and contours) and velocity distributions (arrows) on $x = 0$ plane are plotted for Model DL [($\alpha, \omega, \amt$) = (1, 0.5, 0.2)] at the same epochs of \fig{bar1}.
In each lower panel, $v_{\rm z}$ (false color and contours) and  velocity vectors ($v_y$, $v_z$: arrows) on the $x$=0 plane are plotted.
The contours, arrows, and notations have the same meaning as in \fig{core1}.
}
\label{fig:bar2}
\end{center}
\end{figure}

In this section, we show the third type: the `bar fragmentation' model (model DL).
The cloud with a large initial amplitude of the non-axisymmetric perturbation and either a high rotation speed or strong magnetic field has changed shape from a disk to a bar  in the isothermal collapse phase [see Paper I \S 4.4].
The bar grows from a disk and finally fragments, as shown in Nakamura \& Li (2002, 2003). 
Figs.~\ref{fig:bar1} and \ref{fig:bar2} are the same as Figs.~\ref{fig:core1} and \ref{fig:core2} but for model DL [($\alpha$, $\omega$, $\amt$) = (1.0, 0.5, 0.2)].
See $\S$ 4.4 of Paper I for the isothermal collapse phase of model DL.  
This cloud has an oblateness of 5.2 and an axis ratio of 15.1 at the end of the isothermal collapse phase.
A very thin bar has already formed by the core formation epoch.
Four panels of Figs.~\ref{fig:bar1} and \ref{fig:bar2} correspond to snapshots at $t_c=$ 773 yr (a), 1033 yr (b), 1133 yr (c) and 1235 yr (d).
In model DL,  fragmentation occurs immediately after the  core formation.

In \fig{bar1} (a), two density enhancements are seen in the bar  near  ($x$, $y$) $\simeq$ ($20$ AU, $35$ AU) and ($-20$ AU, $-35$ AU).
This panel shows that gas flows into the bar mainly perpendicular to the major axis of the bar on the $z = 0$ plane, while inside the bar, gas flows to the center along the major axis of the bar.
\fig{bar2} indicates a cut by the $x = 0$ plane which is almost perpendicular to the major axis of the bar.
\fig{bar2} (a) shows that upwardly moving gas is already distributed above the bar.

In Figs. \ref{fig:bar1}(b) and \ref{fig:bar2}(b), the density of the fragment at ($x$, $y$) $\simeq$ ($15$ AU, $32$ AU) and ($-15$ AU, $-32$ AU) is increased up to $5.9 \times 10^{12} \cm$, and the fragments are gradually transformed into a sphere from an elongated bar.
On the other hand, the central density of the bar $n_c$ decreases from  $2\times 10^{11} \cm $ [\fig{bar1} (a)] to $6 \times 10^{10} \cm$ [\fig{bar1} (b)].
This indicates that  gas flows into fragments even from the center of the bar.
The fragments, which are located near ($x$, $y$) $\simeq$ ($16$ AU, $30$ AU) and ($-16$AU, $-30$AU) in \fig{bar1}(c), are more dense and compact than those in \fig{bar1} (b).
At this stage, the central part of the bar is squeezed and narrowed.

In \fig{bar1} (d), the density at the center decreases to $3\times 10^{10} \cm$, and the former bar-shaped adiabatic core is now separated into two parts, while the maximum density of the fragments reaches $ \sim 10^{14} \cm$.
At this stage, the fragments are located at ($x$, $y$) $\simeq$ ($10$ AU, $25$ AU), and ($-10$ AU, $-25$ AU), having moved from  ($16$ AU, $30$ AU), and ($-16$ AU, $-30$ AU) of \fig{bar1} (c).
This means that the fragments are gradually approaching each other as they contract.

The lower panels of \fig{bar2}(a) and (b) indicate that large-scale outflows are ejected both from the front and rear of the bar.
This outflow has a velocity of 0.52 km\,s$^{-1}$ at maximum, while the maximum infall velocity is equal to 1.41 km\,s$^{-1}$.
These figures show that the bar is enclosed in this outflow gas and the top of the outflow reaches $\simeq$ 130 AU at this time.

The lower panel of \fig{bar2}(c) clearly shows that two types of outflows co-exist, that is, one is connected to the bar, and the other is connected to each fragment.
The outer bar-connected outflows are already seen in Fig.~\ref{fig:bar2} (a) and (b). 
On the other hand, the inner fragment-connected outflows seem to be driven by the spin of the fragment cores, and these outflows grow after the fragments contract sufficiently.
The outer outflow reaches $z\simeq$ 200 AU, which has  0.8 km~s$^{-1}$ of the maximum outflow velocity, while the inner one reaches $z\simeq$63 AU with  3.48 km\,s$^{-1}$  of the maximum velocity.
It should be noted that the maximum outflow speed of the inner outflow is much higher than that of the outer.
At this stage, since the infall velocity is as fast as  1.57 km~s$^{-1}$, the maximum outflow speed exceeds the infall one.
\fig{bar2} (c) upper panel shows that gas moves upwardly ($v_z > 0$) for $r \la 50$ AU and a low-density gas ($n_{\rm c} \simeq 3\times 10^8 \cm$) begins to rise toward the adiabatic core.
This is caused by the inner fast outflow, where gas infalls at $t_c = 1033$ yr of \fig{bar1}(b).

A strong mass ejection is seen in \fig{bar2} (d), which is related to the contracting core.
This outflow is faster and denser than the outer outflow.
This is due to the fact that the spin angular rotation speed of the fragments is much higher than that of the rotation of the bar. 
The top of the inner outflow reaches  $z\simeq$ 120 AU, and the flow has a maximum outflow velocity of  5.46 km\,s$^{-1}$.
The fragments have changed shape to  compact spheres from elongated prolate spheroids.
At the same time, the fragments contract and increase their spin speed.
Due to this rapid rotation, magnetic field lines are tightly twisted and fast outflows are ejected.
Since the two lobes of the fast outflow have the same rotational direction, further calculation shows that they seem to merge to form a composite outflow.
On the other hand, the outer outflows survive without merging.

\begin{figure}
\begin{center}
\includegraphics[width=150mm]{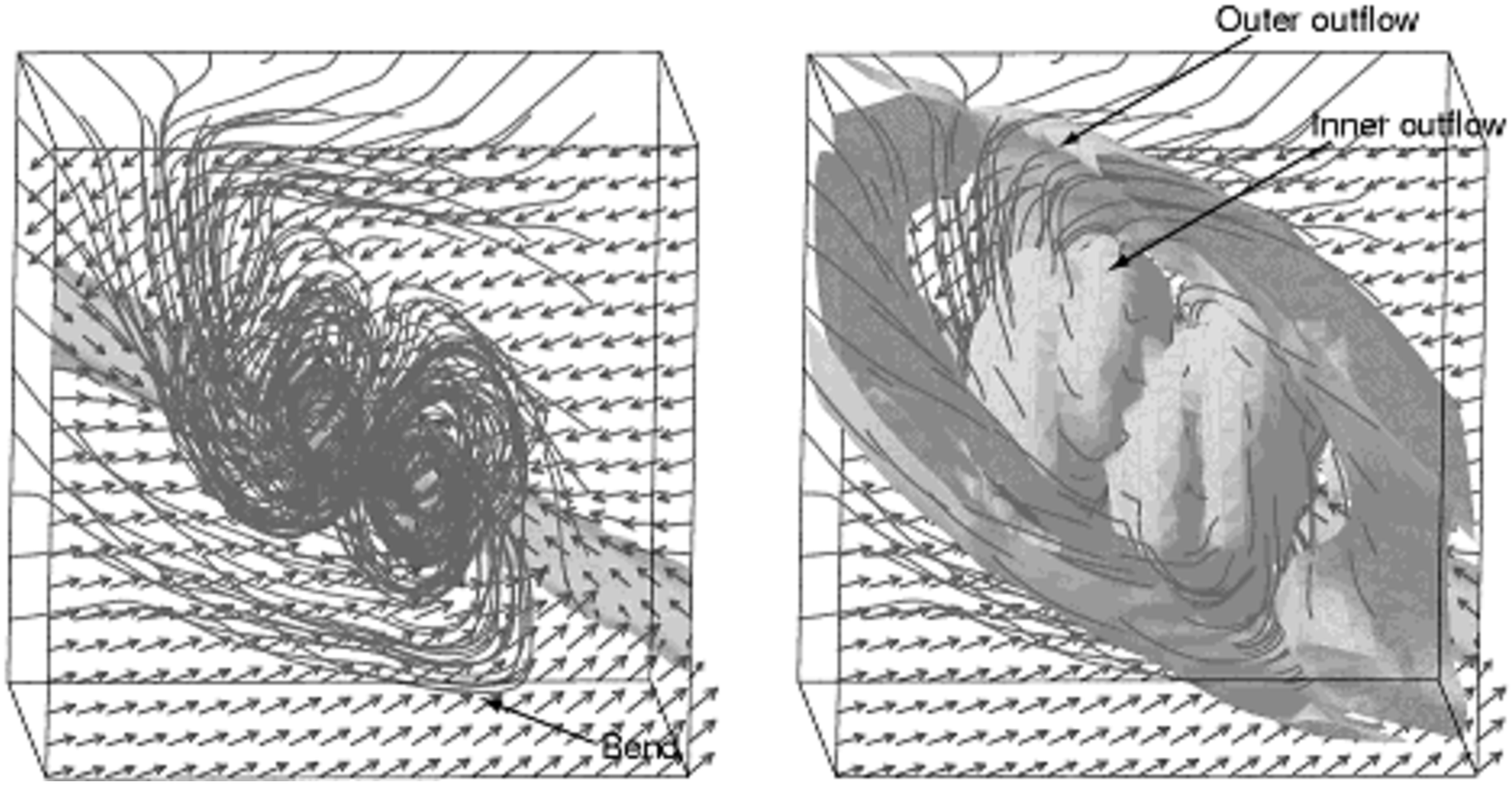}
\caption{
The magnetic field lines and outflow region are plotted for DL [($\alpha, \omega, \amt$) = (1, 0.5, 0.2)] at the same epoch of \fig{bar1} (d).
Magnetic field lines (stream lines), adiabatic core (isodensity surface) and velocity vectors (arrows) on $z$=0 plane are plotted in the left panel.
In the right panel, the outflow regions (isovelocity surface of $v_z = 0$) are also plotted beside the magnetic field lines, the adiabatic core and the velocity vectors.
The box size is 300 AU.
}
\label{fig:bar3}
\end{center}
\end{figure}

\fig{bar3} shows the magnetic field lines (left panel) and the outflow regions (right panel).
This figure shows snapshots of the same epoch of Figs. \ref{fig:bar1}(d) and \ref{fig:bar2}(d), and the box scale is equal to 300 AU.
In the left panel, we find that the magnetic field lines in this region are anchored mainly to the fragments and they are tightly twisted by the spin of the fragments.
When the bar is formed, the magnetic field lines run perpendicularly to the bar and are slightly twisted, as in Fig.~11 of Paper I.
After the fragmentation, the magnetic field lines gather around the fragments as they contract.
At the same time, the fragments increase their spin rotation speed and, as a result, the magnetic field lines are twisted in the region above the fragments.
Conversely, tightly twisted magnetic field lines are evidence of rapid rotation of the fragments.
In the left panel of \fig{bar3}, we can see a bend in the magnetic field lines between the straight outer  and twisted inner field lines.
Gas near this bend coincides with the outflow appearing in the right panel of \fig{bar3}.
Near this bend, the outer outflows which appear in the lower panels of \fig{bar2} are ejected.
These outflows seem to be driven by the rotation of the bar remnant.
The inner two outflows connected to the fragments are seen in the right panels of \fig{bar3}.
It should be noted that the outflow region coincides with the prominent toroidal magnetic fields.

The bar rotates very slowly as seen in Fig.~\ref{fig:bar1} (a) -- (d), while the ring rotates rapidly in the ring fragmentation model [Fig.~\ref{fig:ring1} (a) -- (d)].
Owing to this slow rotation or less efficient centrifugal force, the fragments approach each other gradually in the bar fragmentation case.
Although we have not seen the fragments merge in this model, bar fragmentation models often show mutual merges and form a compound core (see Table~\ref{table:2}).

\section{Discussion}
\label{sec:dis}
\subsection{Cloud Fragmentation}
In this section, we discuss whether the fragmentation occurs or not, depending on the initial magnetic field strength ($\alpha$) and the angular rotation speed ($\omega$).
In order to investigate the fragmentation condition of the molecular cloud, we have studied  144 models with different parameter sets.
Fragmentation occurs in 76 out of the 144 models studied.
Ring fragmentations are seen in 53 of the 76 fragmentation models, while bar fragmentation is seen in 23 models. 
The model parameters and results of all models are summarized in Table~\ref{table:2}.
In this table, the oblateness, axis ratio, central magnetic field strength and angular rotation speed at $n_c = 5\times 10^{10} \cm$ are shown.
The fragmentation mode (the core model, the ring fragmentation, or the bar fragmentation) and the final fate of the fragments, i.e. mergers or survivals, are also described.

\subsubsection{Evolution of the Isothermal Collapse Phase and The Magnetic Flux and Spin ($B$-$\Omega$) Relation}
Although the clouds fragment only in the adiabatic accretion phase ($n_{\rm c} \ge 5 \times 10^{10} \cm$),  the cloud evolutions in the isothermal collapse phase ($n_{\rm c} < 5 \times 10^{10} \cm$) are important in understanding the following fragmentation process in the adiabatic phase. 
The evolutions of the molecular cloud with different initial magnetic field strengths and angular rotation speeds in the isothermal collapse phase are studied in Paper I. 
Here, we summarize these evolutions briefly.

In Paper I, we found the magnetic flux density and the angular rotation speed at the cloud center converge to a relation of \eq{UL} in the process of isothermal collapse, irrespective of the initial amplitude of the non-axisymmetric perturbations.
The first term of \eq{UL} indicates the ratio of rotational to gravitational energy, and the second term means the ratio of the magnetic to thermal energy in the vicinity of the cloud center.
Equation~(\ref{eq:UL}) is plotted with a thick line in \fig{fc}.
We named this curve the magnetic flux-spin ($B$-$\Omega$) relation.
As discussed in Paper I, there are four different evolutional patterns (c.f. models A, B, C and D) according to the initial magnetic flux density and the rotation speed.
Four arrows (models A, B, C and D), which meet the $B$-$\Omega$ relation at the endpoint in \fig{fc}, represent these evolutional paths from the initial state ($n_{\rm 0} \, = \, 5 \times 10^2 \cm$) to the beginning of the adiabatic accretion phase ($n_c= 5\times 10^ {10}\, \cm$).
The evolutions of models B (the core model), C (the ring fragmentation model) and D (the bar fragmentation model) for the adiabatic stage have been described in the preceding section.
Since model A [($\alpha$, $\omega$) = (0.01, 0.01)] forms a compact and dense core and shows evolution similar to that of model B in the adiabatic stage, we abbreviate this evolution in this paper.
Models A-D have different initial amplitudes of the non-axisymmetric perturbation  (for detailed evolution, see \S 4 of Paper I).

Since the $B$-$\Omega$ space is divided into two by the above-mentioned $B$-$\Omega$ relation curve, the evolution of the cloud is also divided into two: that is, in the models distributed in the area below the $B$-$\Omega$ relation curve, the points move towards the upper-right or right, which means the cloud collapses spherically (see \S 4.1 and 4.2 of Paper I).
In this region, the forces supporting the cloud radially (centrifugal and magnetic force) are insufficient  relative to the gravity and the pressure force.
On the other hand, in the models above the $B$-$\Omega$ relation curve, the points move towards the lower-left, which means the cloud collapses vertically making a disk (see \S 4.3 and 4.4 of Paper I).
Thus, we call the region in the lower or left side of the $B$-$\Omega$ relation curve  the `spherical collapse region' and that to its upper or right side as the `vertical collapse region'.
The magnetic flux density and angular rotation speed at the beginning of the adiabatic accretion phase ($n_c = 5 \times 10^{10} \cm$) are concentrated near or on the $B$-$\Omega$ relation curve.
We call models concentrated near the horizontal part of the $B$-$\Omega$ relation curve ($0 < \bb \la 0.3$ and $\ww \simeq 0.2$)  `rotation dominated disks', because the disk is mainly supported by the rotation.
On the other hand, models near the vertical part of the $B$-$\Omega$ relation curve ($\bb \ga 0.3$ and $0 < \ww \la 0.2$) are named  `magnetic dominated disks', because the disk is supported mainly by the magnetic field.

\subsubsection{Fragmentation Condition}
\fig{fc} shows how the cloud  fragments during dynamical contraction.
The types of fragmentation are indicated by large symbols, as $\sq$(ring fragmentation), $\bigtriangleup$(bar fragmentation), $\bigcirc$(ring or bar fragmentation, depending on $\amt$ and $n_{c,0}$), and $\times$(no fragmentation).
The large symbols appear at positions which represent the initial conditions of  $\Omega_c$ and $B_{zc}$.
Models indicated with $\bigcirc$ fragment through the bar mode when the models have large non-axisymmetric perturbations $\amt$, or low initial densities $n_{\rm c,0}$, while they fragment through the ring mode when $\amt$ is small or $n_{\rm c,0}$ is high.
From the distribution of symbols we can divide the parameter space into several regions.
Colored regions indicate the regions where the cloud fragments in the adiabatic accretion phase.
Their colors denote the modes.
Blue, red and violet regions represent, respectively, ring fragmentation, bar fragmentation and either bar or ring fragmentation,  corresponding to the amplitude of $\amt$.

Small symbols, $\diamond$ (fragmentation) and + (no fragmentation), are plotted at the positions which represent the central magnetic field strengths ($\bzc$) and angular rotation speeds ($\omegac$) at the beginning of the adiabatic contraction phase ($ 5 \times 10^{10} \cm$).
Open and filled $\diamond$ symbols, indicate types of fragmentation as ring (open) and bar (filled).
One large symbol ($\bigcirc$, $\sq$ or $\times$) represents several models with the same  magnetic flux density and angular rotation speed, but a different amplitudes of non-axisymmetric perturbation ($\amt=10^{-3},0.01, 0.2$, and  $0.3$), and central density ($n_{c,0}= 5 \times 10^2 \cm, 5 \times 10^4 \cm$ and  $ 5\times 10^6 \cm$).
There are 63 large symbols in this figure, while there are 144 small symbols.

The ring fragmentation occurs in a large area in the initial parameter space shaded blue, while the bar fragmentation occurs when the initial angular velocity is larger than 0.2 [i.e. $\omega \ga 0.2$] irrespective of the initial magnetic flux density.
This is due to the fact that the evolution of the cloud differs greatly between the lower and the upper side of the $B$-$\Omega$ relation curve.
That is, the axis ratio grows slowly in models  below the $B$-$\Omega$ relation curve.
As discussed in MTM04 and Paper I, the non-axisymmetry begins to grow after a disk is formed, namely, growth of the axis ratio depends on the degree of oblateness.
The oblateness grows in proportion to $\rhoc^{1/2}$ in the area above the $B$-$\Omega$ relation curve, while it grows to $\rhoc^{1/6}$ below the curve.
After the oblateness reaches $1$ - $2$, the axis ratio grows in proportion to $\rhoc^{1/6}$ as shown in Paper I.
Therefore, if the initial model parameters are chosen in the lower region, the non-axisymmetry does not grow sufficiently in the isothermal collapse phase because the oblateness grows only slowly.
Even in this region, the non-axisymmetry would grow if the initial cloud density were substantially lower than that of the observed cloud ($n_c \ll 5 \times 10^2 \cm$).
This is because in such a cloud, the isothermal collapse phase would continue for a long time.  
Thus, a cloud with parameters below the $B$-$\Omega$ relation curve does not form an elongated bar.
Therefore, only ring fragmentation occurs for these parameters.

We calculated models with $\amt= 10^{-3}$, 0.01, 0.2, and 0.3 of the initial amplitude of the non-axisymmetric perturbation.
In models in the lower region, no elongated bar appears, even for large non-axisymmetric perturbation $\amt = 0.3$. 
For example, models AL and BL, which have large initial non-axisymmetric perturbations ($\amt=0.2$), only have small axis ratios 0.068 and 0.23 at the end of the isothermal collapse phase ($n_c = 5 \times 10^{10} \cm$).
However, if much larger  non-axisymmetric perturbation (e.g. $\amt > 0.5$) is added to the  model in the lower region, a bar may form in the isothermal collapse phase.
On the other hand, in the model whose parameters exist in the upper side of the curve, an elongated bar appears in the isothermal collapse phase, and the axis ratio grows even in  the  isothermal phase.
Therefore, these models result in either a ring or bar fragmentation, depending on the initial amplitude  of the non-axisymmetric perturbations.
An area in which only bar fragmentation is possible appears in the upper right corner of the figure.
In this region, the ring fragmentation is not possible even if a disk is formed in the isothermal collapse phase.
The disk deforms to a compact core, because the angular momentum of the disk is removed by the magnetic tension.
Neither ring nor bar fragmentation occurs in the unshaded area, in which a small core is formed because the initial rotation speed is slow or  magnetic braking is effective.

\subsubsection{Ring Fragmentation}
\begin{figure}
\begin{center}
\includegraphics[width=140mm]{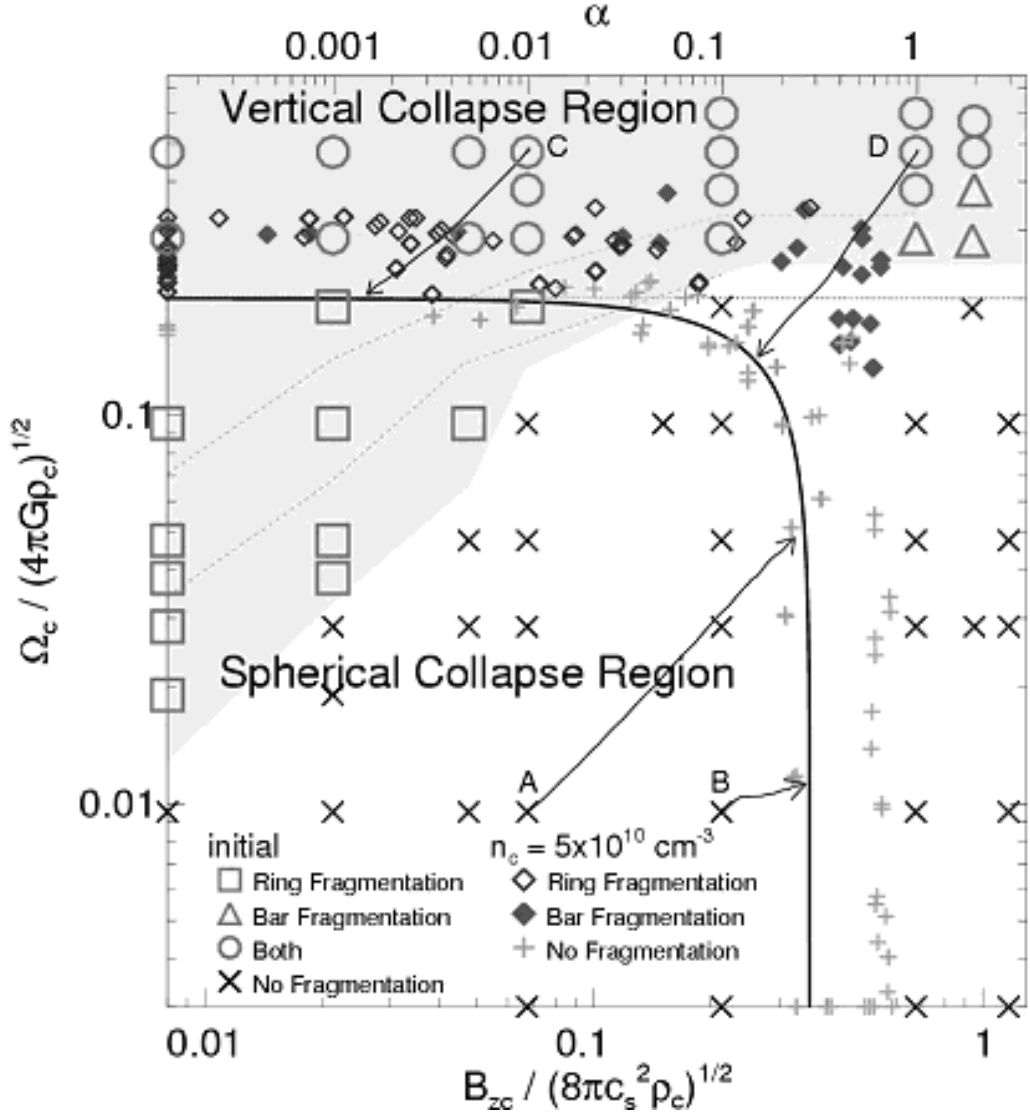}
\caption{
Pattern of fragmentation is shown on the magnetic flux density - rotation speed plane.
The lower axis indicates the square root of the magnetic pressure ($\bzc/\sqrt{8 \pi}$) normalized by the square root of the thermal pressure ($\sqrt{c_s^2 \rhoc}$) and the left ordinate does the angular speed ($\omegac$) normalized by the freefall timescale ($\sqrt{4 \pi {\rm G} \rhoc}$), respectively.
The upper axis indicates the parameter, $\alpha$.
Large symbols mean the fragmentation types: ring $\sq$, bar $\bigtriangleup$, both $\bigcirc$ and no fragmentation $\times$ are plotted against the initial values of $\bzc/(8\pi c_s^2 \rhoc)^{1/2}$ and  $\omegac/(4 \pi  G \rhoc)^{1/2}$.
Small symbols indicate whether fragmentation occurs $\diamond$ (open: ring fragmentation, filled: bar fragmentation) or not +. 
These symbols are plotted against $\bzc/(8\pi c_s^2 \rhoc)^{1/2}$ and  $\omegac/(4 \pi {\rm G} \rhoc)^{1/2}$ at the end of the isothermal phase ($n_{\rm c}= 5 \times 10^{10} \cm$).
Blue, red, and violet  mean the regions of models of the ring, bar, and both fragmentations for the initial parameter space for $n_{\rm c,0} = 5 \times 10^2 \cm$.
The two broken lines indicate the fragmentation region for $n_{\rm c,0} = 5\times 10^4$ and $5 \times 10^5\cm$, respectively.
The dotted line indicates  $\ww=0.2$. 
Four arrows represent the evolutional paths from the initial state ($n_{\rm c,0} \, = \, 5 \times 10^2 \cm$)  for models AS [$(\alpha, \omega, \amt)$ =(0.01, 0.01, 0.01)], BL (0.1, 0.01,  0.2), CS (0.01, 0.5, 0.01) and DL (1, 0.5, 0.2), respectively.
Thick line represents a curve of the magnetic flux - spin relation found in Paper I as $\dfrac{\omegac^2}{(0.2)^2  4\pi G \rho_c} + \dfrac{\bzc^2}{(0.36)^2  8\pi c_s^2 \rhoc } =1 $.
}
\label{fig:fc}
\end{center}
\end{figure}

\begin{figure}
\begin{center}
\includegraphics[width=150mm]{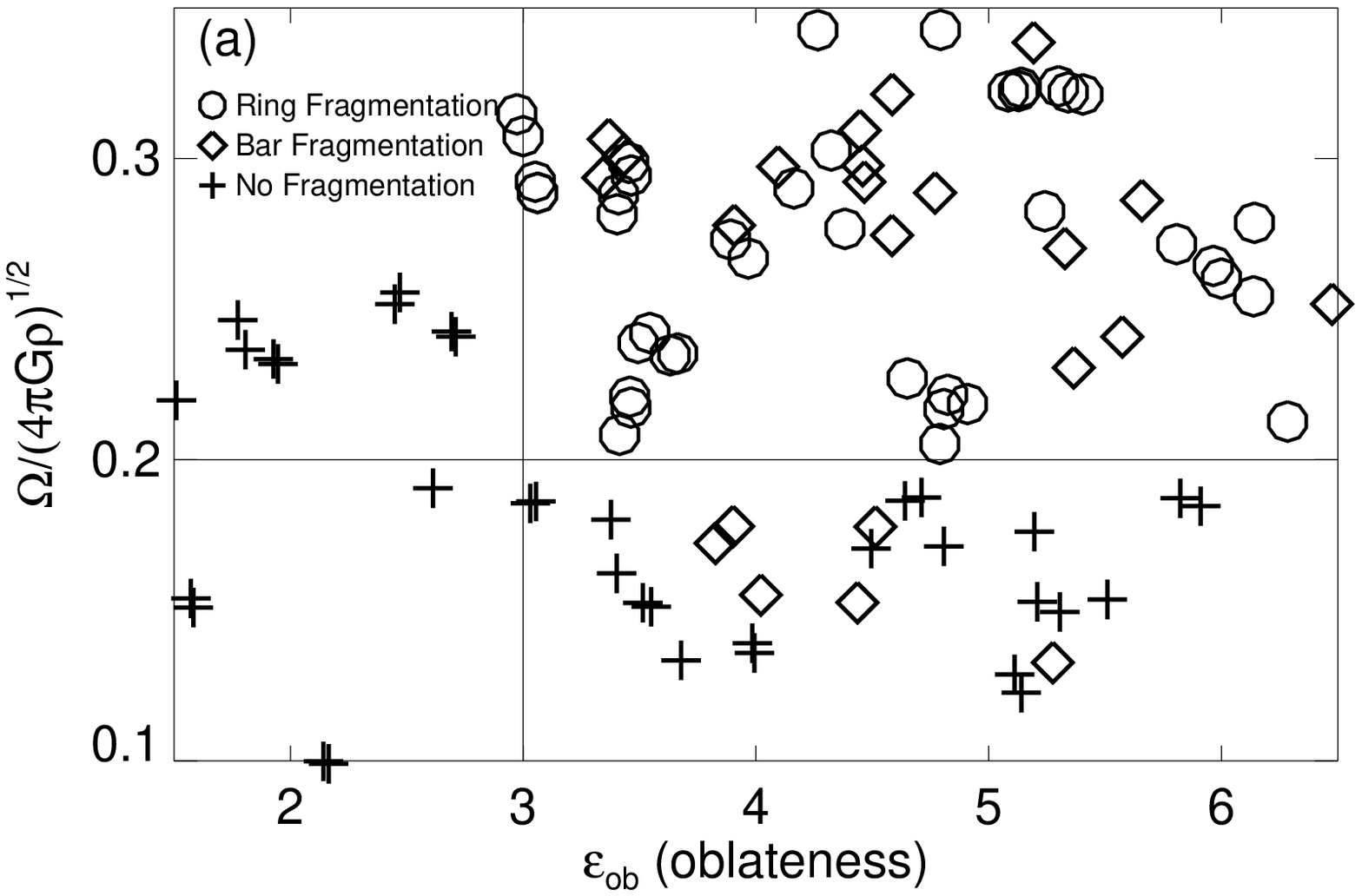}
\includegraphics[width=150mm]{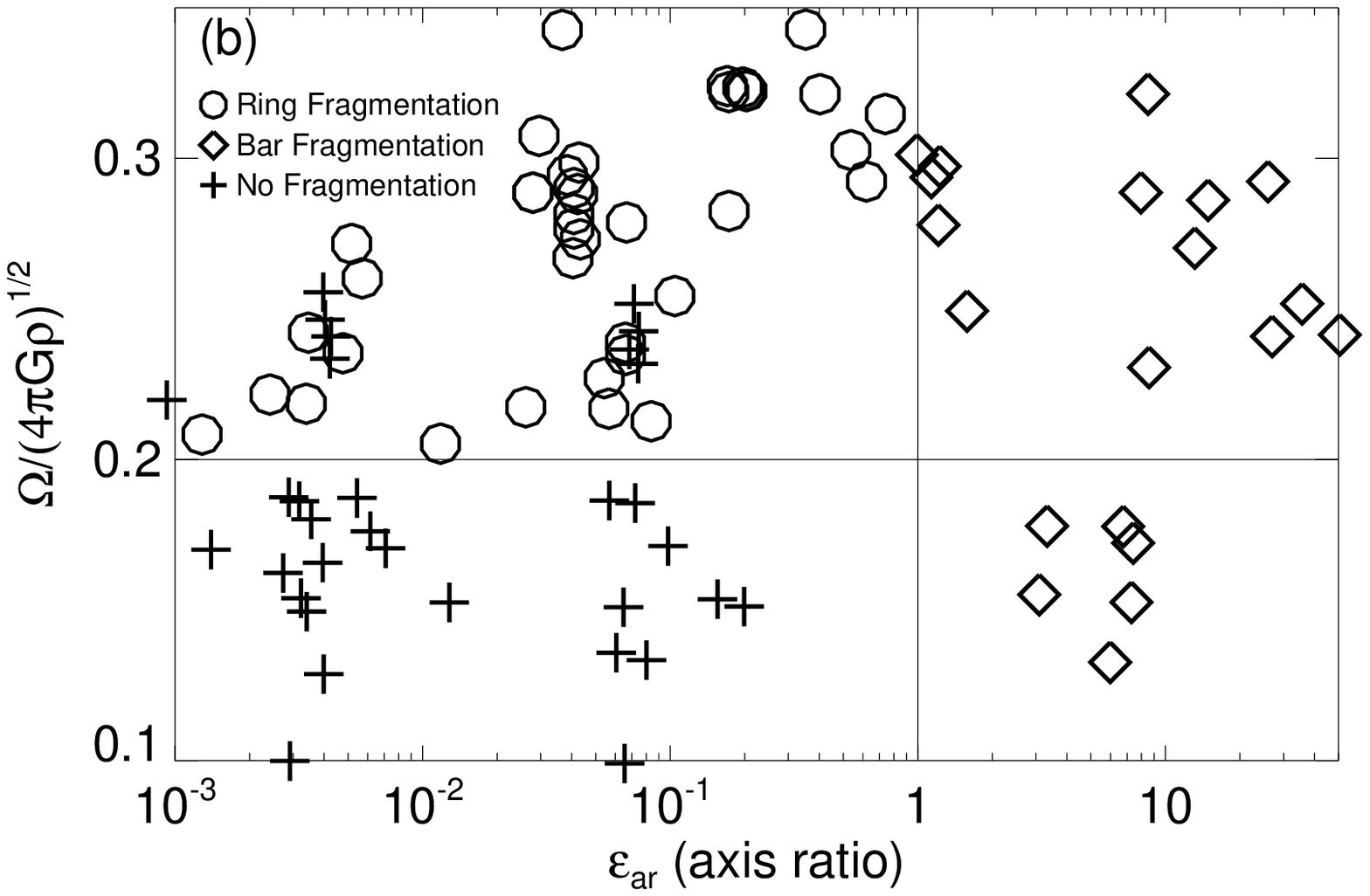}
\caption{
The angular rotation speeds normalized by the free-fall timescale are plotted against the oblateness $\ob$ (upper panel) and the axis ratio $\ar$ (lower panel).
These are values at the beginning of the adiabatic phase ($n_c = 5\times 10^{10} \cm$).
The symbol, $\circ$, $\diamond$, and $+$ represent the ring, bar and no fragmentation, respectively.
}
\label{fig:obar}
\end{center}
\end{figure}

The small open diamond symbols (ring fragmentation), which denote the value of the magnetic field strength and angular velocity at $n_c= 5\times 10^{10} \cm$ in \fig{fc}, are distributed in a narrow band of $0.2 \la \ww \la 0.3$.
This means that whether the core displays the ring fragmentation or not is  determined only by the angular rotation speed at the stage of $n_c \simeq 5 \times 10^{10} \cm$, irrespective of the magnetic field strength (see also \fig{obar}).
If the cloud evolves to have a sufficiently fast rotation speed [$\ww=0.2$] in the isothermal collapse phase, the ring fragmentation occurs in the subsequent adiabatic accretion phase.
As a result, the ring fragmentation condition could be denoted as follows:
\begin{equation}
\ww > 0.2,
\label{eq:fc}
\end{equation}
at the beginning of the adiabatic accretion phase ($n_c = 5 \times 10^{10} \cm$).
This condition is equivalent to that of \citet{matsu03}, in which they did not include the  magnetic field.
If this condition has been satisfied in the isothermal collapse phase, the cloud can fragment through a ring shape as shown in Figs.~\ref{fig:ring1} and \ref{fig:ring2}.
The cloud in the area below the $B$-$\Omega$ relation curve evolves to the upper-right at approximately 45 degrees in \fig{fc}, and the cloud approaches the $B$-$\Omega$ relation curve, as denoted in Paper I.
Clouds with parameters in the upper region continue to evolve to the lower-left until they reach the line.
Therefore only clouds with initial parameters in either blue or violet areas can arrive at $\ww=0.2$ and fragmentation occurs.
In other words, models with large $\alpha$ meet the magnetic-dominated part of the $B$-$\Omega$ relation where $\Omega_c/(4 \pi G \rho_c)^{1/2} < 0.2$.

The oblateness ($\ob$) and the angular velocity [$\ww$] are plotted at the end of the isothermal phase in \fig{obar} (a). 
In this panel, symbols $\circ$, $\diamond$ and $+$ represent the ring, the bar and no fragmentation.
As for ring fragmentation models ($\circ$), the ring fragmentation occurs only in the upper-right area of $\ob > 3$ and $\ww > 0.2$ in \fig{obar} (a).
The ring fragmentation does not occur without a thin disk even if a cloud has $\ww>0.2$ in the isothermal collapse phase.
The models in the upper-left region [$\ob<3$ and $\ww > 0.2$] of \fig{obar}(a) represent the clouds which rotate rapidly and have thick disks or spheres.
These clouds are formed if the initial density is high, because for such clouds, the isothermal contraction phase ends before a sufficient thin disk is formed.
The oblateness at the end of the isothermal phase depends on the initial density, because the oblateness increases with the cloud density as shown in Fig.~3 of Paper I.
Thus, the clouds having extremely high densities ($n_c \ga 10^6 \cm$) initially are distributed in this region.

On the other hand, the clouds in the lower-left region [$\ob<3$ and $\ww < 0.2$] of \fig{obar}(a) do not fragment through the ring mode either, in which these clouds initially have strong magnetic fields.
These clouds form magnetic-dominated disks.
In the lower-left region, the core appears as a thick disk or a sphere rotating slowly and does not fragment.  
As a result, only the rotation dominated-disks in the upper-right region fragment through the ring mode.

Fragmentation depends on the initial density especially for models in the area inside the $B$-$\Omega$ relation curve.
We have plotted the area of the ring fragmentation (i.e. blue shaded area) assuming the initial cloud density to be $5 \times 10^2 \cm$.
However, if we adopt a higher initial density such as $n_{\rm c,0}=5 \times 10^4 \cm$ or $n_{\rm c,0}=5 \times 10^6 \cm$, the lower limit of the angular speed for fragmentation is raised because a only short period of the isothermal collapse phase is available.
In \fig{fc}, three models with different initial densities ($n_{\rm c,0}= 5 \times 10^2 \cm$, $5 \times 10^4 \cm$, and $5 \times 10^6 \cm$) are plotted.
Comparing models with $n_{c,0}= 5\times 10^2 \cm$ and $n_{c,0}= 5 \times 10^4 \cm$, the lower limit of the initial rotation speed for the ring fragmentation increases slightly from  $\omega >0.02$ for $n_{c,0} = 5 \times 10^2 \cm$ to $\omega > 0.04$ for $n_{c,0} = 5 \times 10^4 \cm$ for $\alpha = 0$.
Because the model with a higher initial density does not reach the $B$-$\Omega$ relation curve, the fragmentation condition, \eq{fc}, is not fulfilled in the isothermal collapse phase for higher initial densities.
Thus, although the lower limit of the initial rotation speed for the ring fragmentation depends on the initial density, this effect seems small. 
On the other hand, for models in the upper side of the $B$-$\Omega$ relation curve, their evolutions hardly depend on the initial density.
These clouds evolve rapidly and reach the $B$-$\Omega$ relation curve when the central density increases for two orders of magnitude.
Thus, these clouds induce fragmentation except for extremely high density clouds (for example $n_c \ga 10^8 \cm$).

As for model with  $\alpha = 0$, fragmentation does not occur for $\omega<0.02$.
However, if the initial density is lower than $ 5 \times 10^2 \cm$, or in other words it has a long isothermal collapse phase, fragmentation may occur even for a cloud with $\omega<0.02$.
On the other hand, for $\alpha>0.01$,  clouds with $\omega<0.02$ never fragment, irrespective of their initial densities.
This seems due to the fact that these clouds must evolve into magnetic dominated disks.

\subsubsection{Bar Fragmentation}
Bar fragmentation occurs in the cloud with a large magnetic field and an initially fast rotation speed.
Small filled diamonds indicate the cores of the bar fragmentation in \fig{fc}.
These symbols are almost distributed in the area of  $\ww>0.2$ where the symbols of ring fragmentation (open small diamonds) are also distributed.
These two fragmentation modes co-exist in the same region.
However, several bar fragmentation models are distributed outside the ring fragmentation region.
Bar fragmentation occurs even in the models of strongly magnetized clouds ($\alpha > 1$) if clouds have an initial rotation of $\omega \le 0.3$, while ring fragmentation occurs only in the rapidly rotating and weakly magnetized cloud.
In the bar fragmentation models,  fragmentation occurs in the elongated bar formed in the isothermal collapse phase.
\fig{obar}(b) shows that six bar fragmentation models with angular speeds lower than $\ww < 0.2$ have  large axis ratios $\ar$ = 3.1, 3.3, 6.0, 6.8, 7.2 and 7.5.
They are  special models, that is, models initially in the region of $\Omega_{c,0}/(4 \pi G \rho_{c,0})^{1/2} > 0.2$ and $\alpha \ge 1$.
Since they evolve toward the lower-left in the isothermal phase, $\Omega_c/(4 \pi G \rho_c)^{1/2}$ becomes smaller than 0.2.

Bar fragmentation occurs in the cloud with a large axis ratio, irrespective of the magnetic field strength and the angular speed at the end of the isothermal phase.
For the bar formation, the cloud needs to be located initially above the $B$-$\Omega$ relation curve (i.e. in the vertical collapse region).
The evolution of the axis ratio depends on the initial magnetic field strength and the rotation speed, and the axis ratio obtained at the end of the isothermal phase is an increasing function of $\alpha$ and $\omega$ as shown in Paper I.
However, the cloud which is magnetized strongly or rotating slowly tends to form a small core.
Such a core does not deform to an elongated bar and does not induce fragmentation.
This indicates that the bar fragmentation region in \fig{fc} resembles the ring fragmentation region above the $B$-$\Omega$ relation curve.
In this calculation, we added 30\% of the  non-axisymmetric perturbation at maximum to the initial cloud.
If we add much larger non-axisymmetric perturbations, the lower boundary of the bar fragmentation area (blue area) may extend downward in \fig{fc}.

\subsubsection{Comparison with Previous Work}
In this section, we compare our fragmentation conditions with previous work.
Firstly, we consider the fragmentation conditions of non-magnetized cloud.
In our study, the fragmentation condition is denoted as $\omega > 0.02$ for $n_{\rm c,0} = 5\times 10^2 \cm$ (see \fig{fc}), in a cloud with $\alpha = 0$.
\citet{matsu03} studied the evolution of the non-magnetized rotating cloud with $n= 2.6 \times 10^4 \cm$ of  initial density, and found that fragmentation occurs in the cloud with $\omega > 0.026$, which is mostly in agreement with our conditions for the non-magnetized cloud.

The fragmentation of the magnetized clouds is studied by \citet{boss02} and \citet{hosking04}.
\citet{boss02} assumes the cloud have 200 $\mu$G ($\alpha \simeq 4$) magnetic field strength and $n = 6 \times 10^5 \cm$ initial density, and he adopt the angular rotation speed in the range of $ 10^{-14}$ s$^{-1}$ $-$  $10^{-13}$ s$^{-1}$ ($\omega=0.01-0.12$).
It should be noted that he assumes the magnetic field strength to decrease with time to take into account the effect of the ambipolar diffusion as $B=B_{0,i}(1-t/\tau_{AD}$), where $\tau_{AD}=10 t_{\rm ff}$. 
As a result, he finds that all of the clouds fragment and concludes that the magnetic field promotes the cloud fragmentation.
These clouds are located to the right hand of the $B$-$\Omega$ relation curve in \fig{fc}. 
In our study, the clouds in this region cannot fragment, because these clouds form compact and dense core in the adiabatic stage owing to magnetic braking.
In his study, since magnetic braking is not taken into account, the specific angular momentum of infalling gas seems relatively high.
Thus, only the effect which promotes cloud fragmentation (magnetic pressure and dilution of gravity by magnetic tension) is included in his calculation.
This is consistent with the fact that fragmentation occurs only in the cloud with fast rotation. 

On the other hand, a magnetized (764$\mu$G; $\alpha \simeq 25$) and non-magnetized (0$\mu$G; $\alpha=0$) clouds with the same angular velocity ($\omega\simeq0.2$) and the same density ($n = 2.8 \times 10^6 \cm$) are investigated in \citet{hosking04}.
Note that they included both the ambipolar diffusion and magnetic braking.
We did not include the former while \citet{boss02} did the former in
the approximation form but not the latter. 
The non-magnetized cloud fragments in \citet{hosking04},
while the magnetized cloud does not.
Comparing their parameters with ours, their magnetized cloud ($\alpha = 25$, $\omega = 0.2$) is in the non-fragmentation region in \fig{fc}, while their non-magnetized cloud ($\alpha=0, \omega = 0.2$) is in the fragmentation region.
Therefore their result is consistent with our fragmentation conditions.
We conclude that the magnetic field suppresses the cloud fragmentation.
The opposite result obtained by \citet{boss02} is likely to be 
ascribed to missing magnetic braking.

\subsection{Angular Momentum Redistribution}
In this section, we discuss how the angular momentum is redistributed between the orbital and spin angular momenta in the typical bar and ring fragmentation models.
After fragmentation, the angular momentum of the host cloud is divided into orbital and spin angular momenta of fragments.
As shown in \S \ref{sec:ringf} and \S \ref{sec:barf}, the fragments seem to have large orbital but small spin momenta in ring fragmentation, while they have small orbital and large spin angular momenta in bar fragmentation. 
To estimate the angular momenta of the fragments quantitatively, 
we calculate the total ($j_{\rm tot}$), orbital ($j_{\rm orb}$) and spin ($j_{\rm spn}$) specific angular momenta as follows:
\begin{eqnarray}
j_{\rm tot} & = & j_{\rm orb} + j_{\rm spn},
\label{eq:tot}
\\
j_{\rm orb} &=&  \left( {\bf r}_{\rm f} 
\times {\bf v}_{{\rm f}} \right) \cdot {\bf e}_{\rm z},
\label{eq:orb}
\\
j_{\rm spn } &=& \int^{\,'}_{\rho > \rho_{25} }  \rho({\bf r}) 
\left\{ ( {\bf r} -  {\bf r}_{\rm f}  ) \times [{\bf v({\bf r})}-{\bf v}_{\rm f}] 
 \right\} \ dV \cdot {\bf e}_{\rm z} / M, 
\label{eq:spn}
\\ 
\ M &=& \int^{\,'}_{\rho > \rho_{25}}  \rho({\bf r}) \, dV,
\label{eq:mass}
\end{eqnarray} 
where  
${\bf r}_{\rm f} =\int^{'}_{\rho>\rho_{25}} \rho({\bf r}) \, {\bf r}  \, dV /M$  
($\int^{'}$ represents that summation should be done for one fragment),
${\bf v}_{\rm f} = \int^{'}_{\rho>\rho_{25}} \rho({\bf r}) \, {\bf v}({\bf r}) \, dV/M $, $\rho_{\rm 25}$  and ${\bf e}_{\rm z}$ mean respectively the position vector of the mass center of the fragment, the bulk velocity of the fragment, a typical density of the fragment as $\rho_{25} = 2.5 \times 10^{11} \cm$ and a unit vector of the $z$ direction.
We identify a fragment as a gas which has a density of  $n \geq 2.5 \times 10^{11} \cm$ in equations~(\ref{eq:tot}), (\ref{eq:spn}), and  (\ref{eq:mass}).
Although we have compared how the values depend on the choice of this density, $ 5 \times 10^{12} \cm$, $ 2.5 \times 10^{13} \cm$ and $  5 \times 10^{13} \cm$ give essentially the same result.

\begin{figure}
\begin{center}
\includegraphics[width=180mm]{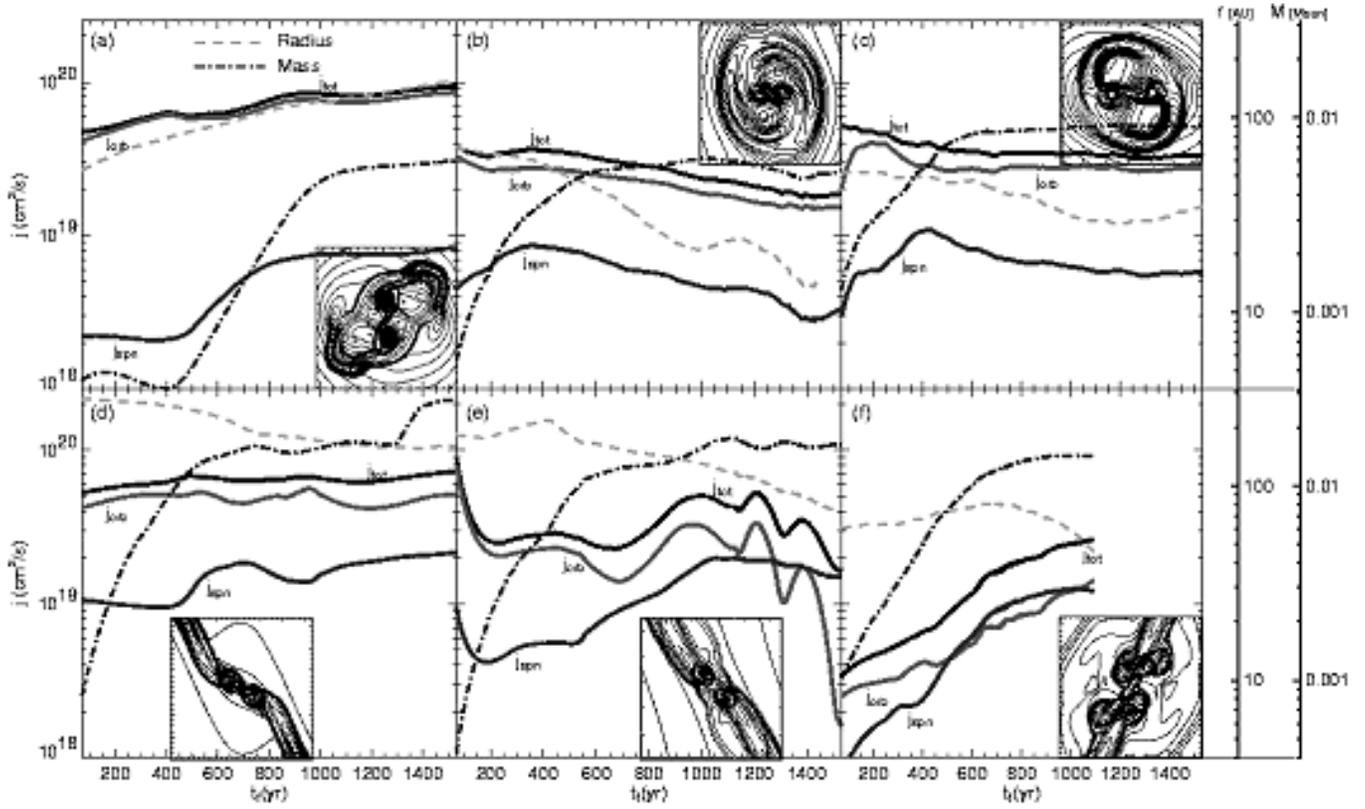}
\caption{
The total ($j_{\rm tot}$), orbital ($j_{\rm orb}$) and spin ($j_{\rm spn}$) specific angular momenta are plotted against the time after fragmentation for  6 models: (a) ($\alpha, \omega, \amt$)= (0.01, 0.5, 0.01), (b) (0.1, 0.6, $10^{-3}$), (c) (1.0, 0.6, $10^{-3}$), (d) (0.01, 0.6, 0.2), (e) (0.1, 0.5, 0.3), and (f) (1.0, 0.5, 0.2).
Distance between each fragment (broken line) and fragments mass (dot-dashed line) is also plotted in each panel and the axes are displayed on the right-hand side of this figure. The upper three panels indicate the ring fragmentation models, while the lower three are the bar fragmentation models.
The contour plot in each panel means the density distribution at about 1200 years after the fragmentation.
}
\label{fig:ang}
\end{center}
\end{figure}

\fig{ang} shows the three specific angular momenta (total, orbital and spin) of the fragments  plotted against the time after fragmentation ($t_f$).
Three ring-fragmentation models (upper panels) and three bar-fragmentation models (lower panels) are plotted in this figure.
Magnetic field strengths increase from the left to the right as  $\alpha$=0.01 [left panels: (a) and (d)], 0.1 [middle panels: (b) and (e)] and 1.0 [right panels: (c) and (f)].
The distances between the fragments are also plotted in each panel.
The contour plots are inset in each panel to show the density distribution of $z$=0 at about 1200 yr after the fragmentation.

\subsubsection{Ring Fragmentation}
The upper panels of \fig{ang} represent the angular momentum distribution in the ring fragmentation models.
From the insets, the ring structure is still seen outside the fragments.
In panel (a), which  corresponds to a model with a weak magnetic field, the fragment has $2.2 \times 10^{18} \jcm$ of the spin specific angular momentum, while it has $4.1 \times 10^{19} \jcm$ of the orbital one at 200 yr after the fragmentation.
The orbital angular momentum is about 20 times larger than the spin angular momentum.
Fragment mass grows up to $5.5 \times 10^{-3} \msun$ in 1400 yr.
At the same time, although the spin angular momentum increases up to $8 \times 10^{18} \jcm$, it is still 10 times smaller than the orbital angular momentum.
The distance between the fragments has increased from 55 AU ($t_f = 200$ yr) to 160 AU ($t_f = 1400$ yr), because of the large angular momentum of the inflow gas.
A similar increase in the spin angular momentum is observed in panels (b) and (c).
It should be noted, in any case of ring fragmentation, the system has a large orbital angular momentum and a small spin momentum.
The ratio of spin to the orbital angular momenta is about 0.1, and this ratio seems to somewhat depend on the magnetic field strength.
We can see from these panels that the total angular momentum is mainly composed of the orbital angular momentum. 
Tightly twisted magnetic field lines surround the ring.
The magnetic field lines anchored to the fragments are weakly twisted as shown in \fig{ring3}. 
In panels (b) and (c), the distances between the fragments decrease or oscillate, while they begin to increase in a further calculation ($t_c>1400$yr).
For all the ring fragmentation models including those in panels (a), (b) and (c), fragments survive without merging.
The distance between fragments decreases in panel (b) and (c), while it continues to increase or keep constant without decreasing in further calculations. 
Therefore, these models seem suitable to form a binary system.

\subsubsection{Bar Fragmentation}
The lower panels of \fig{ang} indicate the angular momentum redistribution in the bar fragmentation models.
The inset contour plots in panels (d), (e) and (f) show that the fragmentation occurs in the bar in these models.
In panel (d) which is the case of a weakly magnetized cloud, the fragment has $1.1 \times 10^{19} \jcm$ of the spin angular momentum, while it has $4.1 \times 10^{19} \jcm$ of the orbital one at 200 yrs after the fragmentation.
In this model, the orbital angular momentum is larger than the spin momentum, similar to the ring fragmentation models.
However, the ratio of orbital to spin angular momenta is not too large compared with that of the ring fragmentation. 
After the fragment mass grows to about $0.02 \msun$ ($t_f = 1000$ yrs), the orbital angular momentum oscillates while the spin momentum continues to increase.
At the same time, the  distance between the fragments decreases gradually.
In this model, two fragments merge at about $10^4$ yrs after the fragmentation epoch.
In the models shown in panel (e), the spin angular momentum continues to increase and catches up with the orbital one at 1260 yrs, and then orbital angular momentum begins to decrease rapidly, while the spin momentum decreases gradually.
In a model with strong magnetic field [panel (f)], the spin angular momentum also catches up with the orbital one at 540 yrs, and then, both momenta increase.
For models shown in panels (d), (e) and (f),  the spin angular momentum exceeds or is equal to the orbital one, when the fragment mass becomes sufficiently larger compared with that of the fragmentation epoch.
Owing to the decrease of the orbital angular momentum, the radial distance between the fragments decreases.
Although the fragments do not merge in the models shown in (e) and (f) in our calculation,  the radial distance decreases steadily.
Thus, the fragments seem to merge, if we continue the calculation further.
Bar fragmentation models have the common feature of the angular momentum being equivalently redistributed into the orbital and spin angular momentum.
Although the bar fragmentation occurs in 23 models in \fig{fc}, the mutual mergers between two fragments occur in  9 models.
However, this number must increase for a longer simulation.
On the other hand, the strong outflows are seen near the fragments in the bar fragmentation models as shown in \fig{bar3}, because of their large spin angular momenta.

In conclusion, we show that the angular momentum redistribution depends strongly on the fragmentation type, ring and bar, although it depends weakly on the magnetic field strength compared with panel (a)-(c) or (d)-(f).

\subsection{Comparison with Observation}

In this section, we apply our fragmentation conditions to several typical molecular clouds in which physical parameters are relatively well fixed.
We find that the fragmentation region in \fig{fc} is well fitted by an expression using initial magnetic field strength ($B$) and angular rotation speed ($\Omega$) as
\begin{eqnarray}
\nonumber
\frac{\Omega}{\sqrt{4 \pi  G \rho}}
> \left( \frac{B}{3 \sqrt{8 \pi c_s^2 \rho}} \right)^{1/2} \; \; \; \; \; \; \; \;
 \; \; \; \; \; \; \; \; \; \; \; \; \; \; \; \;  \; \; \; \; \; \; \; \; \; \; \; \; \; \; \; \;  \\
+ 0.02\  {\rm log}
\left[
\left(
\frac{\rho}{2 \times 10^{-22}\, {\rm g}\cm} 
\right) 
\left( 
\frac{ 2\times 10^{-13}\, {\rm g}\cm}{\rho_{\rm cri}}
\right)
\right]
,
\end{eqnarray}
where $B$, $\Omega$, $c_s$, $\rho$ and $\rho_{\rm cri}$ denote the magnetic field strength, the angular rotation speed,  the sound speed, the central density of the cloud and critical density, respectively.
If we assume the central density, critical density and temperature to be $\rho = 2\times 10^{-20} {\rm g}\cm$($ n= 5 \times 10^3 \cm $), $\rho_{\rm cri} = 2\times 10^{-13} {\rm g}\cm$($ n= 5 \times 10^{10} \cm $). and $T=10$K of the molecular clouds, the above equation can be rewritten as follows:
\begin{equation}
\Omega > 2.0 \times 10^{-14}  \, {\rm s}^{-1} \left[ 
\left( 
\dfrac{B}{1 \mu {\rm G}}
\right)^{1/2} + 0.25
\right] 
\label{eq:fit}
\end{equation}
Table~\ref{table:3} shows the angular rotation speed (upper 8 clouds) and magnetic flux density (lower 5 clouds) of several molecular clouds observed by \citet{arquilla86}, \citet{goodman90} and \citet{crutcher99} as well as the magnetic flux density and rotation speed necessary for fragmentation expected from equation~(\ref{eq:fit}).
For example, if the molecular cloud L1253 with angular velocity of $\Omega=4.0 \times 10^{-14}$s$^{-1}$ has a magnetic field weaker than $B < $3.1 $\mu$ G, the cloud fragments.
On the other hand, if the molecular cloud L134 with magnetic flux density of $B = 11 \mu$G has  a rotation speed faster than $\Omega > 7.1 \times 10^{-14}$~s$^{-1}$, the cloud fragments in the adiabatic accretion phase and has a possibility of forming a binary system.

\subsection{Effects of the Ambipolar diffusion}
In this subsection we discuss the effects of the ambipolar 
diffusion, which is not taken into account in this study.
As discussed in paper I, the ambipolar diffusion has no
appreciable effects as far as the isothermal collapse phase is concerned. 
However, the ambipolar diffusion plays an important role for high-density 
gases of $ n > 10 ^{11-12} \cm $ \citep[see, e.g., ][and the references
therin]{nakano02}.  The ambipolar diffusion may change the outflows
and fragmentation seen in our models seriously. 

First we examine the effects on the two types of outflows,
i.e., outer and inner outflows, shown in \S 3.2 and 3.3.
The outer outflow emerges from the ring or bar structure,
while the inner one does from the dense adiabatic core.
The effect should not be serious for the outer outflow,
since the density does not reach $ \sim 10 ^{12} \cm $ 
even at the base of the outflow.  
On the other hand, the inner outflow should
weaken substantially since the magnetic field is coupled only
weakly with the gas at the base.  Thus the outer outflow will
dominate if we include the ambipolar diffusion.

Next we consider how the ambipolar diffusion effects the fragmentation.
If the adiabatic core has an appreciable amount of
angular momentum, it will fragment as shown in \S 4.1 and \citet{durisen86}.
On the contrary, the adiabatic core will not fragment
either if the core looses its angular momentum through the outflow 
and the magnetic braking, or if it has a small angular momentum 
at the moment of formation.
The ambipolar diffusion reduces the efficiency
of the outflow and magnetic braking.  Thus it renders to promotion of
fragmentation.  

If the effect of ambipolar diffusion is restricted to the reduction 
in the angular momentum transfer, our conclusion that a cloud of
a large $ \Omega / B $ fragments, is secure.  This is because the angular
momentum transfer is already less efficient in our models showing fragmentation.
The reduction in the angular momentum transfer will not change the
result.  On the other hand, the adiabatic core has only a small amount
of the angular momentum at the moment of formation in the models
showing no fragmentation.  If the angular momentum is small, the fragments
will not survive but merge even if once they are formed.

\section*{Acknowledgments}
We have greatly benefited from discussion with Prof.~ M. Y.~ Fujimoto,  
Prof.~ A.~ Habe and Dr.~K. Saigo.
Numerical calculations were carried out with a Fujitsu VPP5000 
at the Astronomical Data Analysis Center, 
the National Astronomical Observatory of Japan.
This work was supported partially by 
the Grants-in-Aid from MEXT (15340062, 14540233 [KT], 
16740115 [TM]).

\clearpage


\begin{table}
\setlength{\tabcolsep}{4pt}
\caption{Model Parameters for Typical Models}
\label{table:1}
\begin{center}
\begin{tabular}{cccccccccccccc}
\hline
 \multicolumn{2}{c}{Model}  
& $\alpha$  & $\omega$  & $\amz$ & $\amt$& $n_{\rm c,0}$ 
& $B_{\rm zc,0}$ {\scriptsize ($\mu$G)}
& $\Omega_{\rm c,0}$ ({\scriptsize $10^{-8}$ rad \, yr$^{-1}$}) & $M$ {\scriptsize ($\msun$)} 
& $L$ {\scriptsize ($10^5$ AU)} & $M/M_{\rm B,cri}$\\
\hline 
&BL& 0.1 & 0.01& 0.1& 0.2& $5 \times 10^2$ $\cm$ &0.931& 1.26& 12.2& 6.82 & 4.2\\
&CS& 0.01&  0.5& 0.1&0.01& $5 \times 10^2$ $\cm$ &0.295& 63.1& 20.6& 5.94 & 14.4\\
&DL& 1   &  0.5& 0.1& 0.2& $5 \times 10^2$ $\cm$ &2.95 & 63.1& 28.7& 5.71 & 1.4\\
\hline
\end{tabular}
\end{center}
\end{table}

\clearpage
\begin{table}
\vspace{-11mm}
\begin{center}
\caption{Result of All Models}
 \begin{minipage}{140mm}
\setlength{\tabcolsep}{3pt}
\renewcommand{\arraystretch}{0.5}
\vspace{-5mm}
\begin{tabular}{ccccc|ccccccccccccc} \hline \vspace{-2mm}\\ \vspace{-1mm}
 & $\alpha$ & $\omega$ & $\amt$ & $n_{\rm 0}$   & 
$\ob$ & $\ar$    & $B^*$& $\Omega^*$& frag. mode
&survive 
\footnote{The column $\ob$, $\ar$, $B^*$, and $\Omega^*$ are the oblateness, the axis ratio, 
the square root of the magnetic pressure normalized by the thermal pressure [$B_{\rm zc}/(8 \pi c_s^2 \rhoc)^{1/2}$] and the angular speed normalized by the freefall timescale [$\Omega_c/(4 \pi G \rhoc)^{1/2}$] at core formation epoch of $n_c=5\times 10^{10} \cm$, respectively.
The column `frag. mode' represents the fragmentation mode, and `ring',  `bar' and `---' represent  `ring fragmentation model', `bar fragmentation model' and 'no fragmentation model', respectively. 
The column headed by `survive' shows whether the fragments survive or merge after fragmentation as Y: survive to form a binary system, N: merge to form a single protostar.
One row contains two models. That is, result shown in the parenthesis is for the model with initial parameters in the parenthesis. 
For example, the first row indicates that $\varepsilon_{\rm ob}=1.3$ for $A_\phi=0.01$ and
     $\varepsilon_{\rm ob}=1.4$ for $A_\phi=0.2$.
} 
 \\ \hline
1 &0    &0.01&0.01\,(0.2)&$5^2$
\footnote{$5\times 10^2$.}
&1.3\,(1.4)&3.9$^{-3}$\,(6.4$^{-2}$) &0\,(0)      &0.17\,(0.16)&---\,(---)&$-$\,($-$\,)           \\
2 &0    &0.02&0.01\,(0.2)&$5^2$&2.9\,(2.9)&4.0$^{-3}$\,(6.8$^{-2}$) &0\,(0)      &0.25\,(0.24)&ring\,(---)&N\,($-$\,)       \\
3 &0    &0.03&0.01\,(0.2)&$5^2$&3.1\,(3.0)&4.0$^{-3}$\,(7.1$^{-2}$) &0\,(0)      &0.26\,(0.25)&ring\,(ring)&N\,(Y\,)       \\

4 &0    &0.04&0.01\,(0.2)&$5^2$&3.5\,(3.5)&3.5$^{-3}$\,(6.6$^{-2}$) &0\,(0)      &0.24\,(0.24)&ring\,(ring)&Y\,(Y\,)& \\
5 &0    &0.05&0.01\,(0.2)&$5^2$&4.8\,(4.6)&2.4$^{-3}$\,(5.4$^{-2}$) &0\,(0)      &0.22\,(0.23)&ring\,(ring)&Y\,(Y\,)&\\
6 &0    &0.1 &0.01\,(0.2)&$5^2$&5.8\,(6.0)&5.2$^{-3}$\,(1.1$^{-1}$) &0\,(0)      &0.27\,(0.26)&ring\,(ring)&Y\,(Y\,)&\\
7 &0    &0.3 &0.01\,(0.2)&$5^2$&3.5\,(3.4)&4.3$^{-2}$\,(1.2\ \ \ \,)&0\,(0)      &0.30\,(0.31)&ring\,(bar )&Y\,(Y\,)&\\
8 &0    &0.5 &0.01\,(0.2)&$5^2$&5.1\,(4.6)&2.0$^{-1}$\,(9.8\ \ \ \,)&0\,(0)      &0.32\,(0.31)&ring\,(bar )&Y\,(N\,)&\\
9 &0.001&0.01&0.01\,(0.2)&$5^2$&1.6\,(1.6)&3.2$^{-3}$\,(6.4$^{-2}$) &0.23\,(0.22)&0.15\,(0.15)&---\,(---)&$-$ ($-$)             &\\
10&0.001&0.02&0.01\,(0.2)&$5^2$&1.9\,(1.9)&4.2$^{-3}$\,(7.4$^{-2}$) &0.19\,(0.17)&0.23\,(0.23)&---\,(---)&$-$ ($-$)             &\\
11&0.001&0.03&0.01\,(0.2)&$5^2$&2.7\,(2.7)&4.3$^{-3}$\,(7.5$^{-2}$) &0.14\,(0.14)&0.24\,(0.24)&---\,(---)&$-$ ($-$)             &\\
12&0.001&0.04&0.01\,(0.2)&$5^2$&3.7\,(3.6)&4.8$^{-3}$\,(6.6$^{-2}$) &0.10\,(0.10)&0.24\,(0.23)&ring\,(---)&Y\,($-$)     &\\
13&0.001&0.05&0.01\,(0.2)&$5^2$&4.9\,(4.8)&3.4$^{-3}$\,(5.6$^{-2}$) &0.07\,(0.07)&0.22\,(0.22)&ring\,(ring)&Y\,(Y\,)&\\
14&0.001&0.1 &0.01\,(0.2)&$5^2$&6.0\,(6.1)&5.7$^{-3}$\,(1.0$^{-1}$) &0.04\,(0.04)&0.26\,(0.25)&ring\,(ring)&Y\,(Y\,)&\\
15&0.001&0.2 &0.01\,(0.2)&$5^2$&3.0\,(3.0)&3.0$^{-2}$\,(7.4$^{-2}$) &0.03\,(0.03)&0.31\,(0.31)&ring\,(ring)&Y\,(Y\,)&\\
16&0.001&0.3 &0.01\,(0.2)&$5^2$&3.4\,(3.3)&4.2$^{-2}$\,(1.1\ \ \ \,)&0.02\,(0.02)&0.29\,(0.29)&ring\,(bar )&Y\,(N\,)&\\
17&0.001&0.4 &0.01\,(0.2)&$5^2$&5.1\,(3.3)&2.0$^{-1}$\,(1.1\ \ \ \,)&0.01\,(0.02)&0.32\,(0.29)&ring\,(bar )&Y\,(N\,)&\\
18&0.005&0.01&0.01\,(0.2)&$5^2$&2.1\,(2.2)&2.9$^{-3}$\,(6.5$^{-2}$) &0.38\,(0.36)&0.01\,(0.01)&---\,(---)&$-$ ($-$\,)             & \\
19&0.005&0.03&0.01\,(0.2)&$5^2$&3.1\,(3.0)&3.2$^{-3}$\,(7.2$^{-2}$) &0.26\,(0.26)&0.19\,(0.19)&---\,(---)&$-$ ($-$\,)             & \\
20&0.005&0.05&0.01\,(0.2)&$5^2$&4.7\,(4.6)&2.9$^{-3}$\,(5.7$^{-2}$) &0.16\,(0.16)&0.19\,(0.19)&---\,(---)&$-$ ($-$\,)             & \\
21&0.005&0.1 &0.01\,(0.2)&$5^2$&6.3\,(6.3)&4.8$^{-3}$\,(8.4$^{-2}$) &0.08\,(0.08)&0.21\,(0.21)&ring\,(---)&Y\,($-$\,)     & \\
22&0.005&0.3 &0.01\,(0.2)&$5^2$&3.5\,(3.4)&3.8$^{-2}$\,(9.9$^{-1}$) &0.04\,(0.04)&0.29\,(0.30)&ring\,(bar )&Y\,(N\,)& \\
23&0.005&0.5 &0.01\,(0.2)&$5^2$&5.3\,(4.6)&1.7$^{-1}$\,(8.5\ \ \ \,)&0.02\,(0.02)&0.32\,(0.32)&---\,(bar )&$-$\,(N\,)        & \\
24&0.01 &0   &0.01\,(0.2)&$5^2$&2.9\,(2.9)&2.8$^{-3}$\,(7.1$^{-2}$) &0.41\,(0.40)&0\,(0)      &---\,(---)&$-$ ($-$\,)             &\\
25&0.01 &0.01&0.01\,(0.2)&$5^2$&2.9\,(3.0)&3.5$^{-3}$\,(6.8$^{-2}$) &0.38\,(0.38)&0.06\,(0.06)&---\,(---)&$-$ ($-$\,)             &&  \\
26&0.01 &0.03&0.01\,(0.2)&$5^2$&3.7\,(3.7)&4.0$^{-3}$\,(8.0$^{-2}$) &0.29\,(0.29)&0.13\,(0.13)&---\,(---)&$-$ ($-$\,)             & \\
27&0.01 &0.05&0.01\,(0.2)&$5^2$&5.3\,(5.2)&3.4$^{-3}$\,(7.0$^{-2}$) &0.19\,(0.20)&0.15\,(0.15)&---\,(---)&$-$ ($-$\,)             &\\
28&0.01 &0.1 &0.01\,(0.2)&$5^2$&5.8\,(5.9)&5.4$^{-3}$\,(9.1$^{-2}$) &0.12\,(0.12)&0.21\,(0.20)&---\,(---)&$-$ ($-$\,)             &\\
29&0.01 &0.2 &0.01\,(0.2)&$5^2$&3.1\,(3.1)&2.8$^{-2}$\,(2.8$^{-2}$) &0.09\,(0.09)&0.29\,(0.29)&ring\,(ring)&Y\,(Y\,)&\\
30&0.01 &0.3 &0.01\,(0.2)&$5^2$&3.4\,(3.1)&4.1$^{-2}$\,(6.2$^{-1}$) &0.06\,(0.09)&0.28\,(0.29)&ring\,(ring)&Y\,(Y\,)&\\
31&0.01 &0.4 &0.01\,(0.2)&$5^2$&4.2\,(4.1)&4.1$^{-2}$\,(1.2\ \ \ \,)&0.04\,(0.04)&0.29\,(0.30)&ring\,(bar )&Y\,(N\,)&\\
32&0.01 &0.5 &0.01\,(0.2)&$5^2$&5.1\,(4.5)&2.0$^{-1}$\,(10.2\ \ )   &0.03\,(0.03)&0.32\,(0.30)&ring\,(bar )&Y\,(N\,)&\\
33&0.05  &0.1&0.01\,(0.2)&$5^2$&5.1\,(5.1)&4.0$^{-3}$\,(7.1$^{-2}$) &0.24\,(0.25)&0.13\,(0.12)&---\,(---)&$-$ ($-$\,)             &\\
34&0.1  &0   &0.01\,(0.2)&$5^2$&5.3\,(5.4)&1.3$^{-2}$\,(2.3$^{-1}$) &1.75\,(0.42)&0\,(0)      &---\,(---)&$-$ ($-$\,)             &\\
35&0.1  &0.01&0.01\,(0.2)&$5^2$&5.3\,(5.3)&1.1$^{-2}$\,(2.3$^{-1}$) &0.33\,(0.33)&0.01\,(0.01)&---\,(---)&$-$ ($-$\,)             &\\
36&0.1  &0.03&0.01\,(0.2)&$5^2$&5.7\,(5.7)&1.0$^{-2}$\,(2.3$^{-1}$) &0.31\,(0.31)&0.03\,(0.03)&---\,(---)&$-$ ($-$\,)             & \\
37&0.1  &0.05&0.01\,(0.2)&$5^2$&5.1\,(5.2)&9.9$^{-3}$\,(2.7$^{-1}$) &0.32\,(0.32)&0.05\,(0.05)&---\,(---)&$-$ ($-$\,)             &\\
38&0.1  &0.1 &0.01\,(0.2)&$5^2$&4.5\,(4.6)&5.5$^{-3}$\,(1.6$^{-1}$) &0.31\,(0.31)&0.10\,(0.09)&---\,(---)&$-$ ($-$\,)             &\\
39&0.1  &0.2 &0.01\,(0.2)&$5^2$&3.5\,(3.6)&1.3$^{-2}$\,(2.0$^{-1}$) &0.25\,(0.25)&0.17\,(0.17)&---\,(---)&$-$ ($-$\,)            & \\
40&0.1  &0.3 &0.01\,(0.2)&$5^2$&3.5\,(3.7)&2.6$^{-2}$\,(6.4$^{-1}$) &0.18\,(0.19)&0.22\,(0.22)&ring\,(ring)&Y\,($Y$\,)     &\\
41&0.1  &0.4 &0.01\,(0.2)&$5^2$&4.0\,(3.9)&4.1$^{-2}$\,(1.2\ \ \ \,)&0.15\,(0.15)&0.27\,(0.28)&ring\,(bar )&Y\,(N\,)&\\
42&0.1  &0.5 &0.01\,(0.2)&$5^2$&5.2\,(4.6)&1.7$^{-1}$\,(9.4\ \ \ \,)&0.11\,(0.12)&0.28\,(0.28)&ring\,(bar )&Y\,(N\,)&\\
43&1    &0   &0.01\,(0.2)&$5^2$&3.8\,(3.7)&2.8$^{-2}$\,(8.1$^{-1}$) &0.51\,(0.53)&0\,(0)      &---\,(---)&$-$ ($-$\,)             &\\
44&1    &0.01&0.01\,(0.2)&$5^2$&3.7\,(3.7)&2.9$^{-2}$\,(8.2$^{-1}$) &0.54\,(0.53)&0.01\,(0.01)&---\,(---)&$-$ ($-$\,)             &\\
45&1    &0.03&0.01\,(0.2)&$5^2$&3.8\,(3.8)&3.0$^{-2}$\,(7.9$^{-1}$) &0.52\,(0.52)&0.02\,(0.01)&---\,(---)&$-$ ($-$\,)             & \\
46&1    &0.05&0.01\,(0.2)&$5^2$&3.7\,(3.7)&3.4$^{-2}$\,(9.2$^{-1}$) &0.53\,(0.53)&0.03\,(0.02)&---\,(---)&$-$ ($-$\,)             &\\
47&1    &0.1 &0.01\,(0.2)&$5^2$&3.7\,(3.7)&3.6$^{-2}$\,(1.0\ \ \ \,)&0.53\,(0.53)&0.05\,(0.06)&---\,(---)&$-$ ($-$\,)             &\\
48&1    &0.2 &0.01\,(0.2)&$5^2$&4.0\,(4.5)&3.1$^{-2}$\,(3.1\, \, \,)&0.46\,(0.43)&0.18\,(0.15)&---\,(bar )&$-$ ($-$\,)             & \\
49&1    &0.3 &0.01\,(0.2)&$5^2$&3.9\,(4.0)&6.1$^{-2}$\,(6.8\ \ \ \,)&0.46\,(0.46)&0.14\,(0.17)&---\,(bar )&$-$ (N\,)	    &\\
50&1    &0.4 &0.01\,(0.2)&$5^2$&4.8\,(3.8)&9.8$^{-2}$\,(7.5\ \ \ \,)&0.37\,(0.51)&0.21\,(0.17)&ring\,(bar )&Y\,(N\,)&\\
51&1    &0.5 &0.01\,(0.2)&$5^2$&4.8\,(5.2)&3.5$^{-1}$\,(15.1\ \ )   &0.36\,(0.35)&0.34\,(0.34)&ring\,(bar )&Y\,(N\,)&\\
52&2    &0.03&0.01\,(0.2)&$5^2$&5.1\,(5.0)&4.9$^{-2}$\,(1.5\ \ \ \,)&0.55\,(0.56)&0.01\,(0.01)&---\,(---)&$-$\ ($-$\,)            & \\
53&2    &0.4 &0.01\,(0.2)&$5^2$&5.5\,(5.4)&1.6$^{-1}$\,(8.6\ \ \ \,)&0.44\,(0.49)&0.15\,(0.23)&---\,(---)&$-$\ ($-$\,)            &\\
54&2    &0.5 &0.01\,(0.2)&$5^2$&4.3\,(5.6)&5.4$^{-2}$\,(27.0\ \ )   &0.49\,(0.55)&0.30\,(0.24)&ring\,(bar )&Y\,(N\,)&\\
55&2    &0.6 &0.01\,(0.2)&$5^2$&7.7\,(7.3)&1.5\ \ \  \ (50.6\ \ )   &0.30\,(0.44)&0.25\,(0.24)&ring\,(bar )&Y\,(N\,)& \\
56&3    &0   &0.01\,(0.2)&$5^2$&6.7\,(5.9)&3.7$^{-2}$\,(1.6\ \ \ \,)&0.50\,(0.59)&0\,(0)      &---\,(---)&$-$ ($-$\,)             &\\
57&3    &0.01&0.01\,(0.2)&$5^2$&6.0\,(5.9)&5.2$^{-2}$\,(1.6\ \ \ \,)&0.57\,(0.57)&0.01\,(0.01)&---\,(---)&$-$ ($-$\,)             &\\
58&3    &0.03&0.01\,(0.2)&$5^2$&6.4\,(6.2)&4.9$^{-2}$\,(1.5\ \ \ \,)&0.55\,(0.53)&0.01\,(0.01)&---\,(---)&$-$ ($-$\,)             & \\
59&3    &0.05&0.01\,(0.2)&$5^2$&1.4\,(1.3)&1.0$^{-2}$\,(4.7$^{-1}$) &4.03\,(2.50)&0.06\,(0.01)&---\,(---)&$-$ ($-$\,)             & \\
60&3    &0.1 &0.01\,(0.2)&$5^2$&5.9\,(5.8)&6.0$^{-2}$\,(2.0\ \ \ \,)&0.58\,(0.58)&0.03\,(0.03)&---\,(bar )&$-$\,(N\,)&\\
61&1 (2)&0.3 &0.3 &$5^2$&3.9\,(5.2)&3.3\, \, \, (6.0\, \, \,)       &0.46\,(0.52)&0.18\,(0.13)&bar\,(bar ) &N\,(N)    & \\
62&1 (2)&0.4 &0.3 &$5^2$&4.4\,(5.7)&7.2\, \, \, (14.9\, \,)         &0.42\,(0.49)&0.15\,(0.29)&bar\,(bar ) &Y\,(Y)& \\
63&1 (2)&0.5 &0.3&$5^2$&5.8\,(6.5)&21.3\, \, (35.6\, \,)&0.37\,(0.37)&0.27\,(0.25)&bar\,(bar) &Y\,(Y)&  \\
64&0.1\,(1)&0.6 &0.001&$5^2$& 4.2\,(6.2)&3.6$^{-2}$\,(7.9$^{-2}$)  & 0.10\,(0.05) &0.34\,(0.23)&ring\,(ring)&Y\,(Y) & \\
65&0    &0.05&0.01&$5^4$\,($5^6$)&1.7\,(1.2)&5.4$^{-4}$\,(1.4$^{-3}$)&0\,(0)      &0.24\,(0.17)&ring\,(---)&Y\,($-$\,)     &\\
66&0    &0.1 &0.01&$5^4$\,($5^6$)&3.4\,(1.5)&1.3$^{-3}$\,(9.2$^{-4}$)&0\,(0)      &0.21\,(0.21)&ring\,(ring)&Y\,(N\,)&\\
67&0.01 &0.2 &0.01&$5^4$\,($5^6$)&5.2\,(2.6)&6.1$^{-3}$\,(1.6$^{-3}$)&0.05\,(0.06)&0.20\,(0.21)&ring\,(ring)&Y\,(N\,)& \\
68&0.01 &0.3 &0.01&$5^4$\,($5^6$)&4.8\,(3.4)&1.2$^{-2}$\,(3.5$^{-3}$)&0.04\,(0.04)&0.20\,(0.20)&ring\,(ring)&Y\,(N\,)& \\
69&0.01 &0.5 &0.01&$5^4$\,($5^6$)&4.3\,(3.7)&4.1$^{-2}$\,(1.5$^{-2}$)&0.03\,(0.03)&0.28\,(0.23)&ring\,(ring)&Y\,(Y)& \\
70&0.1  &0.3 &0.01&$5^4$\,($5^6$)&4.5\,(3.4)&7.1$^{-3}$\,(2.7$^{-3}$)&0.13\,(0.13)&0.21\,(0.21)&ring\,(ring)&Y\,(Y)& \\
71&0.1  &0.5 &0.01&$5^4$\,($5^6$)&3.9\,(3.6)&4.3$^{-2}$\,(1.6$^{-2}$)&0.12\,(0.10)&0.27\,(0.24)&ring\,(ring)&Y\,(Y)& \\
72&1    &0.5 &0.01&$5^4$\,($5^6$)&3.1\,(5.5)&1.1$^{-2}$\,(1.7$^{-2}$)&0.32\,(0.26)&0.21\,(0.20)&ring\,(ring)&Y\,(Y)& \\
\hline
\end{tabular}
\vspace{-8mm}
\label{table:2}
\end{minipage}
\end{center}
\end{table}

\clearpage

\begin{table}
\caption{
The Magnetic Flux Density and The Rotation Speed Necessary for Fragmentation}
\label{table:3}
\begin{center}
\setlength{\tabcolsep}{3pt}
\begin{tabular}{cccl}
\hline
Object & $\Omega$ {\scriptsize (10$^{-14}$ s$^{-1}$)}  
&  $B$ {\scriptsize ($\mu$G)} & reference \\
\hline 
L1174    & 2.8  & $<$ 1.3 \,& \citet{goodman90}  \\
B35A     & 3.4  & $<$ 2.1 \,& \citet{goodman90}  \\
L1253    & 4.0  & $<$ 3.1 \,& \citet{arquilla86} \\
L1257    & 5.6  & $<$ 6.5 \,& \citet{arquilla86} \\
B163     & 6.6  & $<$ 9.3 \,& \citet{arquilla86} \\
TMC-2A   & 7.1  & $<$ 10.9& \citet{goodman90}    \\
B163SW   & 9.7  & $<$ 21.6& \citet{arquilla86} \\
L1495NW  & 12.6 & $<$ 36.7& \citet{goodman90}  \\
Tau16 & $>$ 5.8 &  7& \citet{crutcher99}  \\
L134  & $>$ 7.1 &  11& \citet{crutcher99}  \\
TauG  & $>$ 8.5 &  16& \citet{crutcher99}  \\
W22B  & $>$ 9.0 &  18& \citet{crutcher99}  \\
W49B  & $>$ 9.7 &  21& \citet{crutcher99}  \\

\hline
\end{tabular}
\end{center}
\end{table}


\clearpage
\clearpage

\clearpage

\clearpage


\clearpage

\clearpage

\clearpage

\clearpage

\clearpage

\clearpage

\clearpage

\clearpage

\clearpage

\end{document}